\documentclass[twocolumn]{aastex7}

\usepackage{amsmath}
\usepackage[dvipsnames]{xcolor}
\usepackage{physics}

\usepackage{hyperref}
\newcommand{\muG}{\,\mu\textrm{G}}
\newcommand{\mG}{\,\textrm{mG}}
\newcommand{\pc}{\,\textrm{pc}}
\newcommand{\cm}{\,\textrm{cm}}
\newcommand{\kyr}{\,\textrm{kyr}}
\newcommand{\Myr}{\,\textrm{Myr}}

\newcommand{\GeV}{\,\textrm{GeV}}

\newcommand{\eVcm}{\,\textrm{eV cm}^{-3}}

\newcommand{\ergcms}{\,\textrm{erg cm}^{-3}\textrm{s}^{-1}}
\newcommand{\ergcmcms}{\,\textrm{erg cm}^{3}\textrm{s}^{-1}}
\newcommand{\cmcms}{\,\textrm{cm}^2\textrm{ s}^{-1}}
\newcommand{\kms}{\,\textrm{km s}^{-1}}

\newcommand{\Alfven}{Alfvén }
\newcommand{\GHz}{\,\textrm{GHz}}
\newcommand{\cellspc}{\,\textrm{cells pc}^{-1}}

\graphicspath{{./}{img/}}
\newcommand{\textmra}[1]{#1}
\newcommand{\mathmra}[1]{#1}

\begin{document}

\title{The Effects of Cosmic Ray Protons on Galactic Nonthermal Filaments}
\author[0009-0004-1105-3244]{Mohan Richter-Addo}
\affiliation{Department of Astronomy, University of Wisconsin-Madison, 475 N. Charter St., Madison, WI 53706, USA}
\email{richteraddo@wisc.edu}
\author[0000-0003-4776-940X]{Roark Habegger}
\affiliation{Department of Astronomy, University of Wisconsin-Madison, 475 N. Charter St., Madison, WI 53706, USA}
\email{rhabegger@wisc.edu}
\author[0000-0003-4821-713X]{Ellen Zweibel}
\affiliation{Department of Astronomy, University of Wisconsin-Madison, 475 N. Charter St., Madison, WI 53706, USA}
\affiliation{Department of Physics, University of Wisconsin-Madison, 1150 University Ave., Madison, WI 53706, USA}
\email{zweibel@astro.wisc.edu}
\author[0000-0002-5811-0136]{Dylan M. Par\'e}
\affiliation{Joint ALMA Observatory, Alonso de Cordova 3107, Vitacura, Casilla 19001, Santiago de Chile, Chile}
\affiliation{National Radio Astronomy Observatory, 520 Edgemont Road, Charlottesville, VA 22903, USA}
\email{dylanpare@gmail.com}
\author[0000-0003-0016-0533]{David T. Chuss}
\affiliation{Department of Physics, Villanova University, 800 E. Lancaster Ave., Villanova, PA 19085, USA}
\email{david.chuss@villanova.edu}

\begin{abstract}
The Galactic Center (GC) contains a collection of filaments that are typically tens of parsecs in length, illuminated by synchrotron radiation from cosmic rays (CR). The origin of these nonthermal filaments (NTFs) is unclear. We aim to distinguish two injection mechanisms: the first mechanism posits that NTFs are fueled either by jets from pulsar wind nebulae and are lepton-dominated; the second mechanism posits that NTFs are fueled by accelerated particles from interstellar shocks and are proton-dominated. We explore these mechanisms using the magnetohydrodynamics (MHD) code \texttt{Athena++}, modified to account for radiative and collisional losses, to simulate CR propagation with lepton and proton CR species. We vary parameters such as magnetic field strength, plasma density, and the CR diffusion coefficient to determine how the range of conditions present in the GC can affect CRs' propagation, heating, plasma flow, and the observed synchrotron emission. We find few observable differences between the proton- and lepton-dominated cases, but comparing the models with observed filament properties motivates consideration of a third formation mechanism: the generation of NTFs arise from intermittent structures in \textmra{GC} turbulence.

\end{abstract}

\keywords{Cosmic Rays (329) - Galactic Center (565) - Magnetohydrodynamical simulations (1966)}
\section{Introduction \label{sec:intro}}

Surrounding the Galactic Center and Sagittarius A$^\ast$ is a twisted torus structure known as the Central Molecular Zone (CMZ) \citep{Molinari2011}. This structure is roughly 8.2 kpc away from us and extends roughly 100--200 pc in radius. This region is rich in dust and molecular gas, and it is much more extreme than the rest of the galaxy: its density, temperature, and CR energy density are larger by 1--3 orders of magnitude than disk material \citep{Heywood2022,Oka2019}.

\begin{figure}
    \centering
    \includegraphics[width=0.95\linewidth]{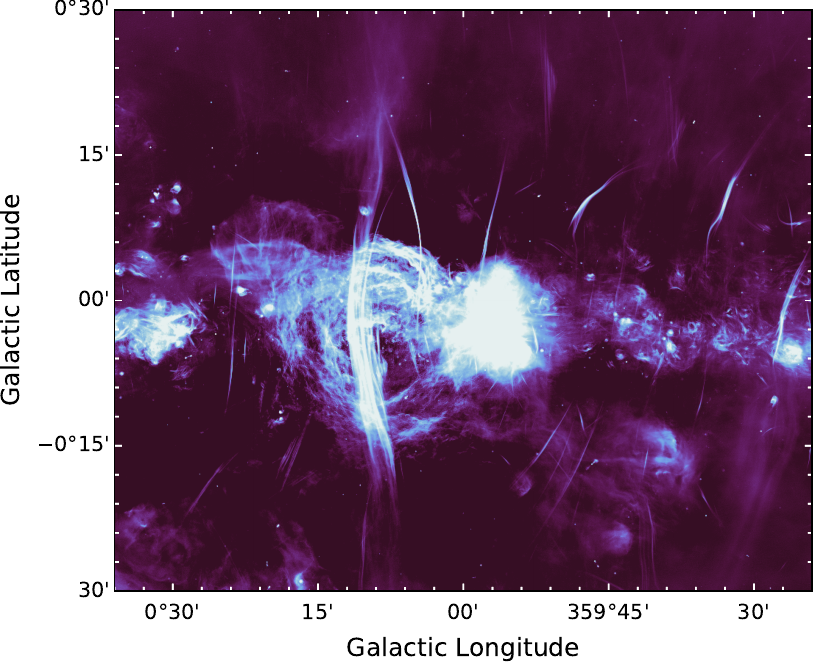}
    \caption{A one square degree image of a portion of the Galactic Plane showing multiple NTFs taken at 1.28 GHz with the MeerKAT telescope, data from \citet{Heywood2022}.}
    \label{fig:LIC}
\end{figure}

Radio observations of the CMZ have detected hundreds of nonthermal synchrotron structures in filament-like formations throughout the CMZ (henceforth referred to as ``nonthermal filaments'' or ``NTFs'') \citep{YZ1984, Morris1985, Bally1989, Nord2004, Heywood2019, Pare2022}. These filaments are preferentially aligned with the magnetic field and suggest that the hot gas contains a predominantly poloidal field centered at the Galactic Center. The magnetic field strength within these NTFs is estimated to be 0.1--1 mG, and the structures extend for tens of parsecs in length, even though they only span less than a tenth of a parsec in width. A one square degree image of a portion of  the Galactic Plane showing multiple NTFs is shown in Figure \ref{fig:LIC}.

The NTFs emit polarized radio waves with a power law, suggestive of nonthermal radiation such as synchrotron emission emitted by cosmic rays (CRs). To emit in this frequency band, these CRs would have Lorentz factors of a few thousand, but the questions of these CRs' origin and their causal connection to the NTFs are still unresolved. Additionally, the poloidal magnetic field topology within the CMZ is still an open question given how many of the filaments appear to not be strictly linear, suggesting interstellar turbulence may be at play \citep{Zhao2025}. The Galactic \textmra{CR} population exhibits a power law specific energy distribution with an index of about -2.7 in the GeV energy range; this corresponds to a synchrotron spectral index of about -0.9, which matches with the statistical mean of NTF observations. This uniformity is not ubiquitous over all filaments, however, as spectral indices between -2 and +0.5 have been observed, and there is also a trend of steepening spectral index in filaments at higher Galactic latitude \citep{Yusef2022a, Pare2022}.

\textmra{Theories about the CRs source include, but are not limited to, magnetic reconnection \citep{Bicknell2001}, magnetized wakes of molecular clouds \citep{Dahlburg2002}, pulsar wind nebulae \citep{Barkov2019}, stellar winds \citep{YusefZadeh2003}, intermittent turbulence \citep{Boldyrev}, and annihilation of light dark matter \citep{Linden2011}.}

\textmra{Our study aims to address scenarios specific to the case of a CR source temporarily passing through the filament. In particular, we aim to} test and differentiate between two particular origin theories: either 1) they are fueled by jets from a pulsar and its surrounding wind nebula, or 2) they are fueled by an interstellar source, such as an interaction between the ambient vertically oriented magnetic field and an azimuthal one in the molecular ring \citep{SM} or a stellar wind termination shock that accelerates interstellar particles \citep{TPE}. The difference between these two origins is that a pulsar wind nebula would inject only electrons and positrons, whereas the interstellar medium (ISM) would inject predominantly protons at a ratio of 100:1 to the leptons \textmra{\citep{YH}, assuming that shock acceleration is more efficient for protons than electrons due to the smaller gyroradii and increased collisional loss rates of electrons relative to protons \citep{Fulat2023, Fulat2026}. While these cases are not comprehensive, we regard them as plausibly prototypical, given the known level of large scale interstellar turbulence and estimated high density of pulsars near the Galactic Center \citep{Federrath2016, Hooper2018}.} This paper aims to address the physical and observational differences between these cases.

The paper is structured as follows: in Section \ref{sec:processes}, we discuss the physical processes that affect CRs in the CMZ. In Section \ref{sec:methods}, we give an overview of \texttt{Athena++}, describe our additional modules, and lay out the parameter variation plan. In Section \ref{sec:results}, we show the results of our simulations and how parameter variation affects CR propagation. In Section \ref{sec:discussion}, we address the key questions we aimed to study and analyze how our results inform the answers. Lastly, in Section \ref{sec:conclusion}, we summarize the paper and present our plans for further research.

\section{Physical Processes \label{sec:processes}}

\subsection{Propagation: Diffusion and Streaming}

\textmra{CRs} spread by scattering off of \Alfven waves, which are generated either by extrinsic turbulence or by the \textmra{CRs} themselves \citep{Zweibel2017, Skilling1975a, Owen2023}. In the case of extrinsic turbulence, the generated \Alfven waves scatter the \textmra{CRs} with a diffusion coefficient $\kappa$ that depends on the properties of the turbulence and transfers momentum between the \textmra{CRs} and the thermal gas. In the case of self-generated waves, an instability is excited if the \textmra{CRs} have a mean drift that is faster than the wave, and this instability slows down the CR drift, such that the CRs convect with the waves\footnote{More precisely, the equilibrium CR drift adjusts such that it is just large enough to overcome damping of the waves by the background medium.}. Momentum is transferred from the \textmra{CRs} to the waves, and the energy that gets removed from the \textmra{CRs} due to wave excitation is removed through dissipation into the background medium \citep{Skilling1975b}. Since waves are always present, diffusion always occurs at some level: the viability of self-confinement under one particular model is assessed in Section \ref{sec:selfconfinement}.

Diffusion allows CRs to disperse independently, resulting in a Gaussian-like distribution over time, and streaming forces CRs to follow an \Alfven group velocity motion, resulting in a plateau-spread of CR energy density with a sharp drop at the edges. These two phenomena are the dominant modes through which CRs spread in a 1D simulation, and they are respectively set by the diffusion coefficient $\kappa$ and the \Alfven speed $v_A=B/\sqrt{4\pi m_in_i}$. The combination of these two propagation methods, combined with both a non-instantaneous injection and tunable parameters, results in a energy profile that is difficult to analytically solve.

We compare these two propagation methods by choosing a characteristic length $l \sim \kappa/v_A$ where diffusion and streaming are codominant. For an \Alfven speed of $44 \kms$---which follows from the fiducial parameters we adopt in this work; see Section \ref{sec:choice}---and a representative length $l=20$ pc, this corresponds to $\kappa\sim2.7 \times 10^{26}\cmcms$, a value assumed in other simulation works such as \citet{TPE}. \textmra{This value, suppressed by around two orders of magnitude from that inferred for the ISM at large \citep{Evoli2019}, aligns with observations that the CR diffusion coefficient is lower in regions of strong MHD turbulence and enhanced magnetic fields \citep{Abeysekara2017,Yang2023,Ng2026}.} The plateaued, uniform surface brightness nature of the filament observations suggests that a lower diffusion coefficient may be necessary to explain the luminosity profile.

\subsection{Loss mechanisms \label{sec:loss}}

We consider synchrotron, inverse Compton, Coulomb, and bremsstrahlung losses by the cosmic-ray electrons, as given in \citet{RL,DM,Schlickeiser2002}. We assume stellar IR photons dominate the inverse Compton losses, whose energy density is scaled by Table 1 of \citet{YH}.

The loss timescales $\tau_i \equiv \gamma/\dot{\gamma}_i$ due to process $i$ for leptons with Lorentz factor $\gamma$ are transcribed in Equations (\ref{eq:loss1})--(\ref{eq:loss2}). We choose to scale the timescales with a frequency $\nu$ centered in the bands observed by \citet{Pare2022}, and we relate this to the CR energy through the synchrotron critical frequency assuming a fixed magnetic field of $B = 200\muG$.

\begin{subequations}\label{eq:losstimes}
\begin{align}
    \tau_{synch} = 0.176 \Myr&\left(\frac{\nu}{8\GHz}\right)^{-1/2}\left(\frac{B}{200\muG}\right)^{-3/2}\label{eq:loss1}\\
    \tau_{bremm} = 2.18 \Myr&\left(\frac{n_p}{100\cm^{-3}}\right)^{-1}\\\nonumber
    \tau_{comp} = 1.76 \Myr&\left(\frac{\nu}{8\GHz}\right)^{-1/2}\left(\frac{B}{200\muG}\right)^{1/2}\\
    &\times\left(\frac{U_{ph}}{100 \,\eVcm}\right)^{-1}\\\nonumber
    \tau_{coul} = 0.949\Myr&\left(\frac{\nu}{8\GHz}\right)^{1/2}\left(\frac{B}{200\muG}\right)^{-1/2}\\
    &\times\left(\frac{n_p}{100\cm^{-3}}\right)^{-1}  \\
    \tau_{tot}|_{\vec{x}=\vec{x_0}} &\equiv \left(\Sigma_i \tau_i^{-1}\right)^{-1} = 0.129\Myr\label{eq:loss2}
\end{align}
\end{subequations}

In selecting these loss processes, we have assumed the thermal gas is fully ionized.  To account for neutral hydrogen we would add ionization losses, which are generally subdominant with $\tau_i\sim 1.73 (100/n_H)$ Myr for $\gamma=3500$ \citep{Ginzburg1964}. \textmra{Perpendicular diffusion does not significantly contribute as a loss process, as from pitch angle scattering, $\kappa_\perp/\kappa_\parallel = (\pi/2)^2(\delta B/B)^4$ \citep{Desiati2014}. In our assumptions of a mostly straight filament, $(\delta{B}/B)^2\sim10^{-3}$, which results in a perpendicular diffusion coefficient 6 orders of magnitude below the parallel diffusion coefficient (See Appendix \ref{sec:selfconfinement}).}

\begin{figure}
    \centering
    \includegraphics[width=\linewidth]{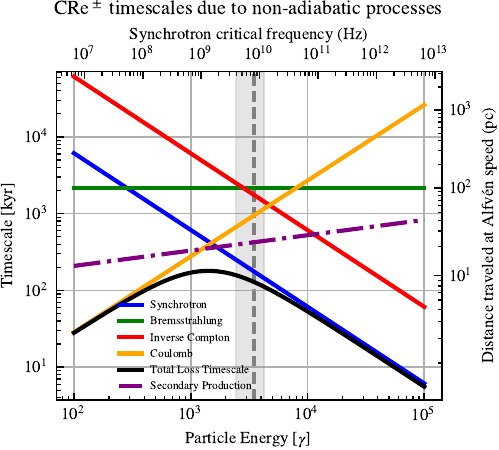}
    \caption{Illustration of the loss timescales \textmra{from Equations (\ref{eq:loss1}--\ref{eq:loss2})} of CRe$^{\pm}$ through various processes as a function of the particle's $\gamma$. The blue line is the synchrotron timescale, the green line is the bremsstrahlung timescale, the red line is the inverse Compton timescale, and the orange line is the Coulomb timescale. The solid black line is the total loss timescale due to these combined \textmra{aforementioned} processes. \textmra{The purple dashdotted line is the timescale associated with secondary production from Equation (\ref{eq:secondary_production}), and isn't included in the total loss timescale.} Both the top and right axes assume a magnetic field of $200\muG$, and the calculated lines additionally assume $n_p=100 \cm^{-3}$ and $U_{ph} = 100 \eVcm$. The gray region denotes the frequency range of 4--12 GHz as observed in the VLA C- and X- bands and the dotted gray line is at the median 8 GHz we assume \citep{Pare2022}.}
    \label{fig:timescales}
\end{figure}

From Equations (\ref{eq:loss1})--(\ref{eq:loss2}), we find not only a total loss time for the particles, but we also note that, for the fiducial parameters, synchrotron radiation is the dominant form of energy loss. These timescales are plotted in Figure \ref{fig:timescales}.

\subsection{Secondary Production \label{sec:sec}}

\textmra{CR} nuclei will inelastically interact with ambient protons in the ISM, creating a shower of secondary particles, such as pions. These secondaries can produce relativistic electrons and positrons. In this way, the existence of \textmra{CR} protons enables the \textmra{CR} leptons to be repopulated. These secondaries then undergo the same loss processes described in Section \ref{sec:loss}.

From \citet{DolagEnsslin}, we find that primary \textmra{CR} protons will produce \textmra{a spectrum of secondary CR leptons with a peak at 1/24th the original CR proton energy $E_c$,}
\begin{equation}\label{eq:q}
    q(\mathmra{E_c}) \approx 64\sigma_{pp}cn_p\frac{C_p}{\GeV}\left( \frac{24\mathmra{E_c}}{\GeV} \right)^{-\frac{4}{3}(\alpha_p-\frac{1}{2})}\,,
\end{equation}
where $C_p$ is the \textmra{normalization} factor present in the \textmra{CRp} spectrum\textmra{, $f_{p}(\mathmra{E_c}) =(C_p/\textrm{GeV})(\mathmra{E_c}/\textrm{GeV})^{-\alpha_p}dE_c$}. Injected CRp would be roughly 100 times more populous than CRe, $C_p=100C_e$, assuming they have the same power law distribution with $\alpha_p=2.6$. \textmra{The inelastic p-p cross-section is $\sigma_{pp}=32\textrm{\,mbarn}$. Treating the cross-section as energy-independent is a conservative treatment that underestimates the importance of secondary production by about 30\% and is highly dependent on the CR spectral index. Since we later argue that the simulated NTFs are much longer compared to the observations and that decreasing the secondary production timescale would cause the lengths of the simulated NTFs to extend even longer, we adopt this larger timescale as a conservative estimate  \citep{SchlickeiserMannheim, Koldobskiy2021}.}

We can calculate a timescale of energy repopulation $\tau_{sec}=f_e(\mathmra{E_c})/q(\mathmra{E_c})$, which indicates the time it would take for the CR energy density to double solely through secondary production:
\begin{equation}
    \tau_{sec}(\mathmra{E_c}) = \frac{f_e(\mathmra{E_c})}{q(\mathmra{E_c})} = \frac{24^{\frac{4}{3}(\alpha-\frac{1}{2})}}{64\sigma_pc}\frac{C_e}{C_p}\frac{1}{n_p}\mathmra{E_c}^{\frac{1}{3}(\alpha_p-2)}\,.
\end{equation}

We are primarily interested in comparing our results to observations of the filaments, so we fix the frequency range to be around 8 GHz. Then, assuming the primary emission is synchrotron, we can convert our equipartition magnetic field to a characteristic CR energy $\mathmra{E_c}$ and redefine our secondary production timescale as
\begin{eqnarray}\label{eq:secondary_production}
    \tau_{sec} = 0.424 \Myr\left(\frac{n_p}{100 \cm^3}\right)^{-1}\left(\frac{\nu}{8 GHz}\right)^{\frac{1}{6}(\alpha_p-2)}\nonumber\\
    \times\left(\frac{B}{200 \muG}\right)^{-\frac{1}{6}(\alpha_p-2)}\,,
\end{eqnarray}
about 3 times longer than the loss time for fiducial parameters. This suggests that production of secondaries can offset radiative and collisional losses only if $n$ is a few hundred cm$^{-3}$ or more\textmra{, larger than the fiducial model (See Table \ref{tab:Simulations_run}).}

\section{Methods \label{sec:methods}}

\subsection{1D MHD equations with CRs in {\normalfont\texttt{Athena++}}}

We model our nonthermal filament in 1D using \texttt{Athena++}, a publicly available astrophysical MHD code written in C++ \citep{Stone2020}. We envisage a situation in which a uniform background medium with constant gas pressure $P_g$, \textmra{CR} pressure $P_c$, and magnetic field $B$ is perturbed by injecting CR energy onto a short interval along a small patch of field lines. We take the injection volume as the midpoint of the filament. We show in Appendix \ref{sec:adiabatic} that for the parameters considered here, the radial expansion of the tube due to \textmra{CR} overpressure is small, justifying the 1D approximation. Using the CR module created by \citet{JiangOh}, we model the CR energy as a \textmra{spectrally unresolved} secondary fluid, and \texttt{Athena++} evolves the two fluids in parallel, tracking their interactions. For 1D simulations, the governing equations are
\begin{subequations}
\begin{equation}\label{eq:mhd1}
    \frac{\partial \rho}{\partial t} + \frac{d\rho u}{dz} = 0\,,
\end{equation}
\begin{equation}\label{eq:mhd2}
   \frac{\partial\rho u}{dt} + \frac{d}{dz}\left( \rho u^2+P_g \right) = \sigma_c\left(F_c-\frac{4}{3}\mathmra{\varepsilon_c} u\right)\,,
\end{equation}
\begin{multline}\label{eq:mhd3}
    \frac{\partial \mathmra{\varepsilon}}{\partial t} + \frac{d}{dz}\left( \left( \mathmra{\varepsilon} + P_g \right)u \right)-\frac{B^2}{8\pi}\frac{du}{dz} \\
    = (u+v_s) \sigma_c\left( F_c - \frac{4}{3}\mathmra{\varepsilon_c} u \right),
\end{multline}
\begin{equation}\label{eq:mhd4}
    \frac{\partial B}{\partial t}=0\,,
\end{equation}
\begin{equation}\label{eq:mhd5}
    \frac{\partial \mathmra{\varepsilon_c}}{\partial t} + \frac{dF_c}{dz} = -(u+v_s) \sigma_c\left( F_c - \frac{4}{3}\mathmra{\varepsilon_c} u \right),
\end{equation}
\begin{equation}\label{eq:mhd6}
    \frac{1}{V_m^2}\frac{\partial F_c}{\partial t} + \frac{1}{3}\frac{d\mathmra{\varepsilon_c}}{dz} = -\sigma_c\left( F_c - \frac{4}{3}\mathmra{\varepsilon_c} u \right)\,.
\end{equation}
\end{subequations}

Here, $\rho$ and $u$ are the thermal gas mass density, and velocity respectively, and
\begin{equation}\label{eq:E}
\mathmra{\varepsilon} = P_g/(\gamma_g-1)+\rho u^2/2+B^2/8\pi
\end{equation}
is the thermal gas energy density, where $P_g$ and $\gamma_g$ are the gas's pressure and adiabatic index respectively. The CRs are described by an energy density $\mathmra{\varepsilon_c}$, flux $F_c$, and a \textmra{CR} transport coefficient $\sigma_c$, which can be related to the diffusion coefficient $\kappa_c$ by Equation (\ref{eq:cr_trans}), with $v_s$ being the \textmra{CR} streaming velocity, which we set equal to the \Alfven velocity $v_A$. $V_m$ is the speed of light parameter, which can be reduced from $c$ itself to reduce computational costs so long as the underlying hydrodynamics remain unchanged. However, we take $V_m$ to equal $c$ itself. The transport coefficient is defined as
\begin{equation}\label{eq:cr_trans}
    \sigma_c^{-1}=\kappa_c+\frac{v_a}{|\nabla P|}(\mathmra{\varepsilon_c}+P_c)\,.
\end{equation}

We compile \texttt{Athena++} with the Roe flux solver, 4 ghost boundary cells, and an adiabatic equation of state ($\gamma_g = 5/3$). We run each simulation with a Courant-Friedrich-Lewy (CFL) number of 0.2, the Runge-Kutta 4th order time integrator, and the Piecewise Parabolic Method for spatial reconstruction \citep{Stone2020}. The simulation boundaries are encoded to be outflowing. We have $2560$ cells over a $40$pc length, resulting in a resolution of $64\cellspc$. Convergence tests are detailed in Appendix \ref{sec:convergence}.

We simulate the injection of CR energy by a traveling object---such as a pulsar or traveling shock---by gradually adding energy to the midpoint of the filament until a total energy $E^\pm$ has been injected when the traveling object (assumed to be moving at $\sim100 \kms$) leaves the filament of diameter $\sim0.2 \pc$.

\subsection{Encoding of Losses and Secondary Production}

To encode how the energy of \textmra{CRs} changes according to secondary production and synchrotron losses (as discussed in Sections \ref{sec:loss} and \ref{sec:sec}), we follow an operator splitting procedure and introduce a simple loss term,
\begin{equation}
    \left. \pdv{\mathmra{\varepsilon_c}}{t} \right|_\mathrm{loss} 
    = - \frac{\mathmra{\varepsilon_c}}{\tau}.
\end{equation}

The timescale $\tau$ is defined as the parallel sum of the loss timescales and the secondary production timescale as defined in Equations (\ref{eq:loss2})\&(\ref{eq:secondary_production}):
\begin{equation}\label{eq:tau}
    \frac{1}{\tau}=\frac{-\mathmra{\varepsilon_c}/dt}{\mathmra{\varepsilon_c}}=\frac{\mathmra{\dot \varepsilon_{c,loss}}-\mathmra{\dot \varepsilon_{c,sec}}}{\mathmra{\varepsilon_c}}=\frac{1}{\tau_{tot}}-\frac{1}{\tau_{sec}}\,.
\end{equation}

During each sub-cycle of the code's time integrator, this term is applied to each cell based on that cell's local properties (in this case, only cosmic-ray energy density $\mathmra{\varepsilon_c}$). The operator splitting approach and simple loss term (first order) treatment is adequate because the timesteps the code takes to handle the \textmra{CR} module (set by the modified speed of light $V_m$) are much smaller than the timescales of these additional processes by six orders of magnitude.

\subsection{Choice of parameters\label{sec:choice}}

\begin{table}[t]
    \footnotesize
    \begin{center}
    \begin{tabular}{|c||c|c|c|c|}
        \hline
        Name of Run & $B [\muG]$ & $\kappa [\cmcms]$ & $n[\cm^{-3}]$ & $E^\pm [\textrm{ergs}]$\\
        \hline
        \hline
        \hline
        Fiducial & 200 & 1e26 & 100 & 5e42\\
        \hline
        \hline
        Bvar\_Econst1 & 50 & 1e26 & 100 & 5e42 \\
        \hline
        Bvar\_Econst2 & 100 & 1e26 & 100 & 5e42 \\
        \hline
        Bvar\_Econst4 & 400 & 1e26 & 100 & 5e42 \\
        \hline
        Bvar\_Econst5 & 800 & 1e26 & 100 & 5e42 \\
        \hline
        \hline
        Bvar\_Lconst1 & 50 & 1e26 & 100 & 6.1e43 \\
        \hline
        Bvar\_Lconst2 & 100 & 1e26 & 100 & 1.7e43 \\
        \hline
        Bvar\_Lconst4 & 400 & 1e26 & 100 & 1.4e42 \\
        \hline
        Bvar\_Lconst5 & 800 & 1e26 & 100 & 4.1e41 \\
        \hline
        \hline
        kvar1 & 200 & 1e25 & 100 & 5e42 \\
        \hline
        kvar2 & 200 & 3e25 & 100 & 5e42 \\
        \hline
        kvar4 & 200 & 3e26 & 100 & 5e42 \\
        \hline
        kvar5 & 200 & 1e27 & 100 & 5e42 \\
        \hline
        \hline
        nvar1 & 200 & 1e26 & 10 & 5e42 \\
        \hline
        nvar2 & 200 & 1e26 & 30 & 5e42 \\
        \hline
        nvar4 & 200 & 1e26 & 300 & 5e42 \\
        \hline
        nvar5 & 200 & 1e26 & 1000 & 5e42 \\
        \hline
        \hline
        hyper\_Econst & 800 & 1e26 & 1000 & 5e42 \\
        \hline
        hyper\_Lconst & 800 & 1e26 & 1000 & 4.1e41 \\
        \hline
        \hline
    \end{tabular}
    \end{center}
    \caption{Tabulation of the different simulations. We use each set of parameters for two simulations each; one with leptons only and one with proton and lepton \textmra{CRs}, multiplying the total \textmra{CR} injection rate by 100. Runs with protons are referred to by appending ``\_p'' to their name.
    We vary the magnetic field in two different ways: in ``Bvar\_\textmra{Econst}[n]'', we vary $B$ while keeping the total energy of injection the same, to emulate the case where the filament changes its properties while the injection method stays the same; in ``Bvar\_Lconst[n]'', we vary both $B$ and $E^{\pm}$ such that the total observed luminosity would stay the same between all five simulations (neglecting losses; see Appendix \ref{sec:luminosityderivation}). The [n=3] line of each variation always has the same parameters as \textmra{``Fiducial,''} so it is not included explicitly.}
    \label{tab:Simulations_run}
\end{table}

Choosing parameters will change the results of the simulation in a few predictable ways. The combination of the magnetic field strength $B$ and the thermal ion density $n_p$ determines the streaming speed $v_A = B/\sqrt{4\pi m_p n_p}$.
Changing $B$ and $n_p$ will affect the plasma beta $\beta = P_g/P_B$, which in turn affects the behavior of the thermal gas after the injection occurs, potentially leading to observable Doppler shifts due to CR-induced pressure flows.   

Guided by data from \citet{Pare2022} and \citet{TPE}, we set our fiducial parameters for a filament to be $B = 200\muG$, $l=20\pc, r=0.1\pc,n_p=100\cm^{-3}, T=10^4\textrm{K}, \kappa = 1\times10^{26}\cmcms$. We also \textmra{consider} a background CR energy density of $1\eVcm$\textmra{, since the solar neighborhood value approximates molecular cloud observations \citep{Aharonian2020}. Assuming a different value would only alter the contrast in Figures \ref{fig:EcLc}d--f.} We vary several of these parameters alongside the total lepton energy of injection $E^\pm$ as in Table \ref{tab:Simulations_run}. We simulate each set of parameters twice; once with leptons only and one with proton \textmra{CRs}, effectively multiplying the total \textmra{CR} injection rate by 100. Runs which include protons are distinguished by a suffix ``\_p''. We note that the energy required to include protons, while model dependent, is generally not untenable. From Table 1, we see that model ``Bvar\_Lconst1\_p'' would require the most energy---$6.1\times 10^{45}$ ergs. For comparison, this is almost exactly the turbulent kinetic energy in one cubic parsec of interstellar gas with $v=10\kms$.

We note that our assumed temperature of $T=10^4\textrm{K}$ for the filaments is higher than expected in the thermal CMZ \citep{Oka2019}. We must assume at least this value to keep the ionization fraction within the NTFs high. If the ionization fraction were low, the effective \Alfven speed for \textmra{CR} propagation would be set by the \textit{plasma} mass density, and therefore $\sim$ 10--100 times larger than what we assume, and the waves from which \textmra{CRs} scatter would be heavily damped, resulting in a larger diffusion coefficient. Both effects would make \textmra{CR} transport much faster. As we we discuss in Section \ref{sec:discussion_length}, the observed filament lengths are hard to explain even with our assumed rate of transport; faster transport would make them even harder to explain.

\begin{figure}
    \centering
    \includegraphics[width=\linewidth]{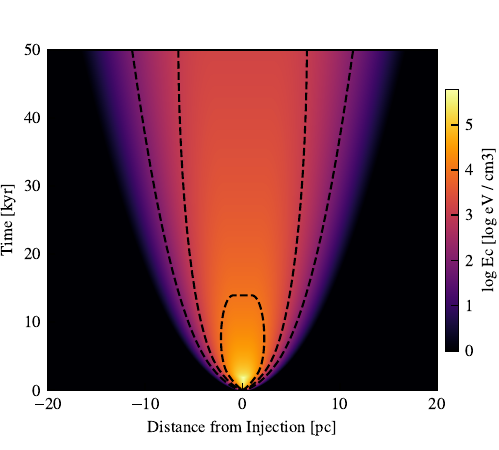}
    \caption{Time evolution of the CR energy density in a 1D flux tube with $\kappa=10^{26}\cmcms$, $B=200\muG$, and $n_e=100\cm^{-3}$, in the case of a proton injection. We see a strong peak persist at the injection site for the duration of the injection which then diffuses and streams outward to overall double the background CR energy density at around 10kyr. Dotted contours trace lines of constant energy density, and are placed at \{$10^2,10^3,10^4$\} $\eVcm$.}
    \label{fig:color}
\end{figure}

\section{Results \label{sec:results}}

\subsection{Spatial Energy and Luminosity Profiles}

\begin{figure*}
    \centering
    \includegraphics[width=\linewidth]{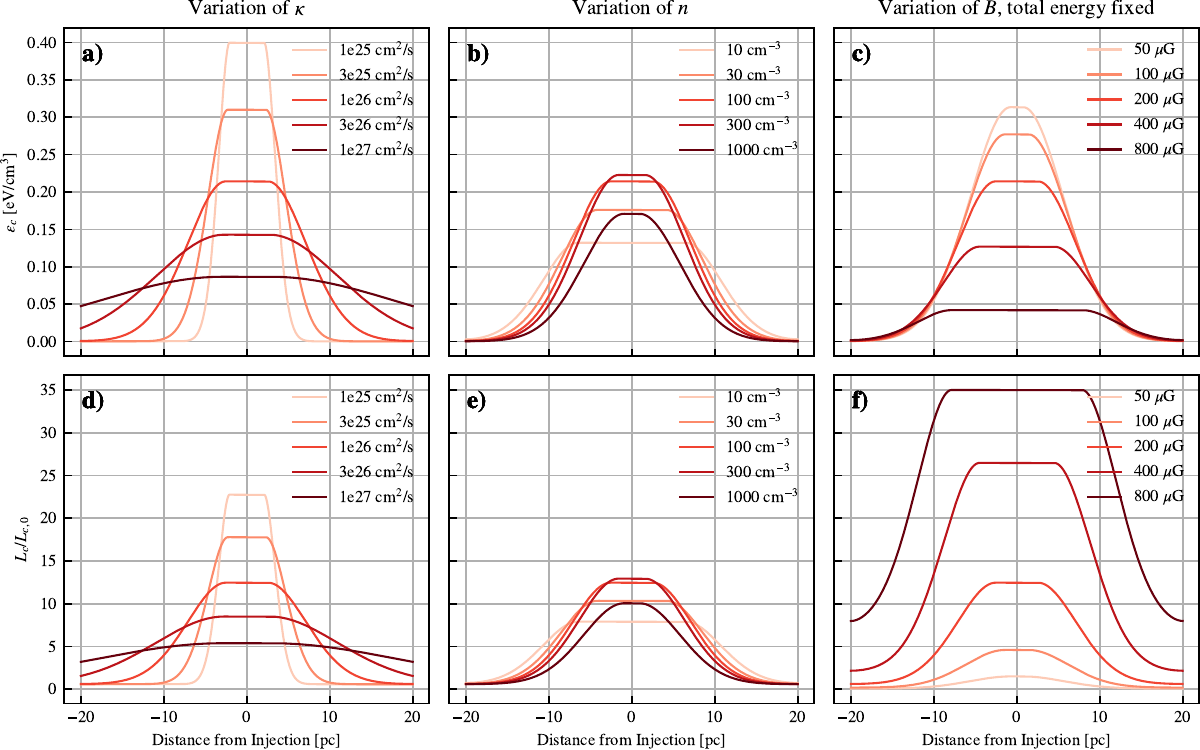}
    \caption{Plots at $t=40 \kyr$ of the energy density (top row) and normalized luminosity per length (bottom row) within these lepton-only simulations. Left column, simulations ``kvar[n]'': When changing the diffusion coefficient, the Gaussian wings get larger while the streaming length stays the same. Middle column, simulations ``nvar[n]'': When changing the number density, the streaming length changes (as well as increases Coulomb losses), while the Gaussian wings stay the same. Right column, simulations ``Bvar\_Econst[n]'': When changing the magnetic field, the streaming length changes and synchrotron losses increase, but here the luminosity is noticeably different as a result of forcing the critical synchrotron frequency to be constant, thus letting the luminosity probe lower-energy segments of the CR spectra. The $\mathmra{\varepsilon_c}$ plots for proton-included injections are roughly the same as these lepton-only simulations, though scaled up by approximately 100x. In parameter variations, variables not mentioned are held constant at the fiducial values.}
    \label{fig:EcLc}
\end{figure*}

Each simulation produces a series of 1D profiles over time. The \textmra{CR} energy density profile as a function of space and time is plotted in Figure \ref{fig:color} as a color plot, arranged such that each horizontal slice is a filament at a different age. As time progresses, the CRs diffuse and stream away from the injection point. At 2 kyr, the injection ends, and the strong peak in the center dissipates. At 40 kyr, the energy profile of the filament appears to have Gaussian wings with a plateaued peak. The details of these effects change when varying certain parameters. In Section \ref{sec:FWHMfitting} we discuss how the profiles provide constraints on $v_A$ and $\kappa$.

Figure \ref{fig:EcLc} shows lepton-only injection plots of three chosen variations and how the energy density as well as observable luminosity would change with the parameters. Varying $\kappa$, as in the left column, changes the shape of the Gaussian wings, but the overall size of the plateau generated by streaming stays approximately the same. Varying $n$ and $B$, as in the middle and right column respectively, changes the streaming velocity $v_A$ without changing the diffusion coefficient $\kappa$. Additionally, in these cases, synchrotron and Coulomb losses will also change, leading to further variations in the energy density and luminosity. The addition of protons would cause the profile to scale up by roughly a factor of 100, with minor variance being attributed to the lack of synchrotron radiation on the protons and the introduction of secondary production.

With a constant magnetic field and spectral index, the synchrotron emissivity per frequency is proportional to the energy density of \textmra{CRs}: this is why the left two columns' luminosity appear to be a scaling of the energy density. For the magnetic field variations, since we hold the observing frequency constant for our observables, different sections of the CR spectrum contribute to the synchrotron emissivity. Following the derivation in Appendix \ref{sec:luminosityderivation}, we find
\begin{equation}
    \frac{L_\nu}{L_0} = \left(\frac{\varepsilon_c}{\varepsilon_{c,0}}+1\right)\left(\frac{B}{B_0}\right)^{(1+\alpha)/2}\,,
\end{equation}
where $B_0$ is the fiducial value of $200\muG$ and $\mathmra{\varepsilon_{c,0}}$ is the background CR energy density of $1\eVcm$. This is the luminosity enhancement due to the \textmra{CR} injection, where a value of 1 for this fraction corresponds to the conditions where this part of the filament is indistinguishable from the background emissivity. In the bottom row of Figure \ref{fig:EcLc}, we plot this value over all variations to determine how observable our simulated filaments are with radio arrays.

\begin{figure*}
    \centering
    \includegraphics[width=\linewidth]{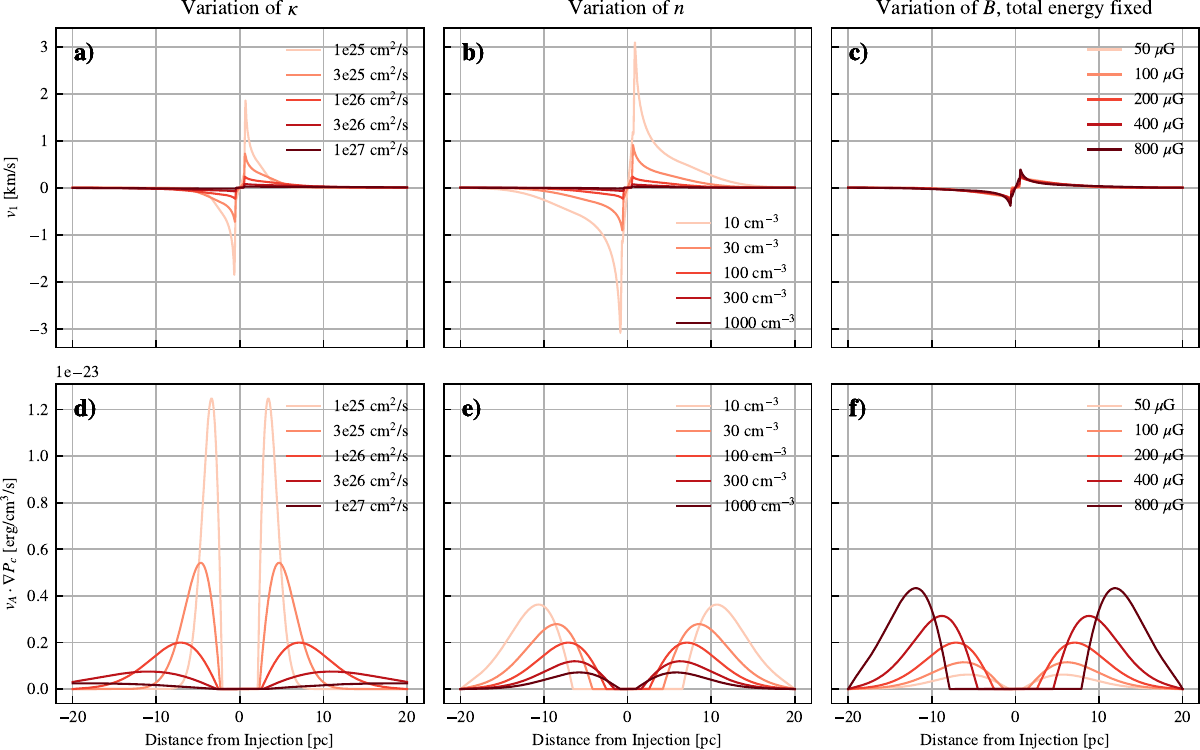}
    \caption{Plots at $t=40 \kyr$ of the induced velocities and heating cause by streaming in the same simulations as Figure \ref{fig:EcLc}, but now including protons. Left column, simulations ``kvar[n]\_p'': When changing the diffusion coefficient, the energy gradient is sharper, which allows for more heating and induced flows. Middle column, simulations ``nvar[n]\_p'': As the number density lowers, the thermal pressure is lessened, which induces more flow in the background; the rise in heating comes from the dependence of $v_A$ on $n$. Right column, simulations ``Bvar\_Econst[n]\_p'': Changing the magnetic field does not affect the induced flow, but it does change $v_A$ which causes the heating to rise. The plots for lepton-only injections are roughly the same as these proton-included simulations, though scaled down by approximately 100x. In parameter variations, variables not mentioned are held constant at the fiducial values.}
    \label{fig:vel1heating_p}
\end{figure*}

\subsection{Induced Flows and Heating\label{sec:heating}}

When CRs are injected into a background ISM component, the increase in CR pressure will push the gas away from the injection point and cause an induced outward flow. It is possible that this induced flow is observable, either as nonthermal broadening, a bulk Doppler shift, or even as a shock. We plot the velocity of the thermal gas in the top row of Figure \ref{fig:vel1heating_p} and find that the velocities decrease with distance from the injection site and increase with decreasing gas density as expected, but never exceed $\sim 2.5\kms$, even at later timesteps. Although this is comparable to or exceeds the thermal line widths of common metals, the volume and column density of the flowing material is so small that it is unlikely to be detectable.

The streaming instability allows the \textmra{CRs} to collisionlessly heat the ISM background through the continuous energy transfer entailed by excitation of waves, which, in a steady state, must dissipate that energy in the background at the same rate, see e.g. \citep{Owen2023}. This heating rate can be written as $\vert\vec{v}_A\cdot\nabla P_c\vert$ and appears in Equations (\ref{eq:mhd3})--(\ref{eq:mhd5}). The plot of this CR heating term is shown in the bottom row of Figure \ref{fig:vel1heating_p}. All plots in Figure \ref{fig:vel1heating_p} are proton-included injections; the plots with lepton-only injections would be scaled down by two orders of magnitude.

We estimate the importance of collisionless \textmra{CR} heating by comparing it to the radiative cooling rate. Taking the optically thin loss function $\Lambda$ to be $\Lambda/n_en_H\sim 2\times 10^{-24} \ergcmcms$ as in \citet{Draine}, we see that at the fiducial density $n_e=n_p=10^2 \cm^{-3}$, the radiative cooling rate is about 3 orders of magnitude larger than the maximum heating rate with protons. Even in cases at the lowest density in our parameter suite (``nvar1\_p''), the \textmra{CR} heating never becomes comparable to $\Lambda n_en_p$.

We have also evaluated the Coulomb heating rate: from Equation (\ref{eq:losstimes}), we see that the energy loss rate due to Coulomb processes is roughly independent of particle energy, so the heating rate is simply $\mathmra{\varepsilon_c}/\tau_{coul}$. Taking $\mathmra{\varepsilon_c}\sim 200 \eVcm$ for a CR proton injection we find that the Coulomb heating rate is of order $10^{-23} \ergcms$, similar---and similarly insignificant---to what we obtained for collisionless heating.

\subsection{Illustrative Example}

\begin{figure}
    \centering
    \includegraphics[width=0.9\linewidth]{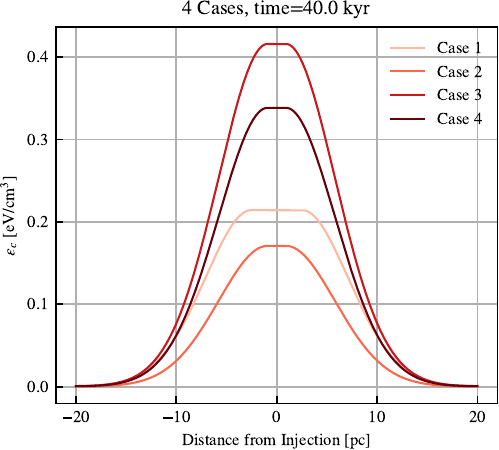}
    \caption{Lepton-only \textmra{CR energy density} profiles of special cases to illustrate the synergistic effects of non-adiabatic losses, secondary production, and protons. Case 1 replots the \textmra{simulation ``Fiducial,''}, Case 2 replots the high density model of ``nvar5,'' Case 3 plots a modified ``nvar5'' model that now includes secondary production, and Case 4 plots a modified '`nvar5'' model while now neglecting non-adiabatic losses. All times are shown 40 kyr after injection.}
    \label{fig:cases}
\end{figure}

To illustrate the detailed interplay between non-adiabatic losses, secondary production, and protons, we create special lepton-only simulations and plot their energy densities 40 kyr after injection in Figure \ref{fig:cases}. Four cases are shown. In Case 1, we replot the \textmra{simulation ``Fiducial.''} In Case 2, we replot the high density case, ``nvar5.'' In Case 3, we plot a variation of ``nvar5'' in which secondary production is included as a source term. In Case 4, we plot another variation of ``nvar5'' but with neither secondary production nor radiative/collisional losses.

The value of Figure \ref{fig:cases} lies in the comparison of energy profiles to Case 1, the fiducial model. Although the \Alfven speed in lower by $\sqrt{10}$ in Case 2, there are more Coulomb losses, so there is less total energy in the simulation. Case 3 keeps the higher Coulomb losses as in Case 2, but we now add secondary production, increasing the energy. Case 4 removes the energy losses and gains that we added in Cases 2 and 3 to show that both effects work to keep the CR energy roughly the same as models without nonadiabatic losses.

\section{Discussion \label{sec:discussion}}

\subsection{Spatial Profile of the filament \label{sec:discussion_length}}

\subsubsection{Constraining the Length of an NTF}

Presumably there is a range of NTF ages in the CMZ, since there is no evidence of some large-scale event that could have mass-produced filaments at a single time. Therefore, it is important to understand what sets their length. We have already seen that unless the filaments are only a few kyr old, the diffusion coefficient $\kappa$ must be significantly smaller than the mean Galactic value. This is also true if propagation is completely diffusive (no streaming) and the length is set by losses; substituting the total loss time of 0.129 Myr from Equation (\ref{eq:loss2}) and setting $\kappa=10^{25} \cmcms$ in the diffusive length scale formula $\sqrt{2\kappa t}$ gives a length of about 3 pc (this would not represent a sharp drop in filament brightness, only a decline by roughly a factor of 2).

Our simulations show that, at 40 kyr after the injection, we have not only created filaments that are 20 pc long, but also they will continue to expand without limit. We should expect to see a lot of filaments longer than 20 pc given our fiducial parameters. Radiative and collisional losses acting alongside secondary production (if protons exist within the injection) will not be able to create a steady-state length, since the protons also diffuse along the magnetic field in the same way as the leptons. Thus, the secondary leptons would be born further away from the center and diffuse more, in a manner akin to that without losses and productions. Therefore, assuming the basic ingredients of our models are the correct ones, the total length of the filament is dependent on its age.

\subsubsection{Characteristics of the Profile \label{sec:FWHMfitting}}

The interaction of streaming and diffusion smears some exact effects, even without considering radiative and collisional losses. Enough characteristics remain to utilize observational fitting and determine parameters of the filament.

\begin{figure}
    \centering
    \includegraphics[width=0.9\linewidth]{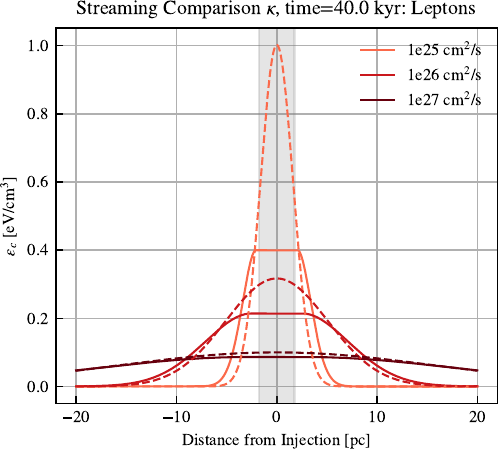}
    \caption{The CR energy density in a lepton-only injection that compares simulations with streaming (solid) and without streaming (dashed). Without streaming, the graph looks remarkably similar to a pure Gaussian. When including streaming, the base shape remains Gaussian but appears plateaued at the peak. The shaded region shows the extent of the plateau in the case of no diffusion $(v_a t)$. Small variation from the Gaussian arise in extreme regions of streaming-dominated and diffusion-dominated regimes. This resulting effect is also seen in cases where the streaming speed increases while the diffusion coefficient is fixed, such as $B$ variation and $n$ variation cases.}
    \label{fig:no_stream}
\end{figure}

The spatial profile of the filament allows for the diffusion coefficient and the \Alfven velocity to be read out. As seen in Figure \ref{fig:no_stream}, the joint diffusion-streaming profile has general agreement with the pure diffusion profile at the Gaussian wings. For the plateau, we find that plotting $v_At$ as the shaded region slightly undershoots the plateau length at streaming-dominated regimes, while at diffusion-dominated regimes, the plateau is barely visible. Therefore, it is possible to fit for and approximate values for $v_At$ and $\sqrt{2\kappa t}$ from observations. Specifically, the standard deviation of the Gaussian wings is $\sqrt{2\kappa t}$ and the width of the plateau is just longer than $2v_At$. 

\begin{figure}
    \centering
    \includegraphics[width=0.9\linewidth]{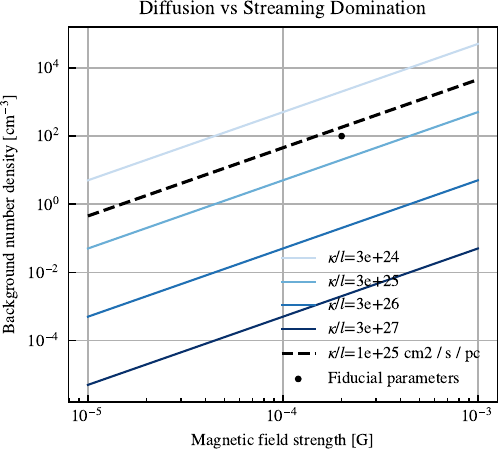}
    \caption{Phase space of diffusion vs streaming domination given specific $B$ and $n_e$ values. Colored lines represent contours where diffusion and streaming have equal contributions: above these lines, diffusion dominates; below these lines, streaming dominates. These lines all follow the analytic form $v_A = \kappa/l$. The colored dot indicates our specific $(B,n_e)$ parameters for our simulation (which roughly corresponds to a $\kappa=10^{26}\cmcms$ if our filament is around 10pc long, unlike our initial choice of $\kappa=3\times10^{28}\cmcms$).}
    \label{fig:diffusionvar}
\end{figure}

Without knowing the exact energy profile, we can also determine which of streaming or diffusion dominates by using the parameters we assume about the region. Equating $t$ within the Gaussian's standard deviation and the half-width of the plateau gives the equation $v_A=\kappa/l$ (where $l$ is the approximate Gaussian FWHM of the filament), the combination of parameters needed to an exactly balanced filament between the two effects. The phase space of this is plotted in Figure \ref{fig:diffusionvar}, where the x-- and y--axis are magnetic field strength and number density respectively. Plotted in color are contours at specific $\kappa/l$ combinations.

The significance of Figure \ref{fig:diffusionvar} lies in the placement of $B$ and $n_e$ parameters with respect to the contours $\kappa/l$. If the choice of $(B,n_e)$ lies below the filament's $\kappa/l$ contour, then that filament is expected to be streaming-dominated, and vice versa.

\subsubsection{Spectral Index Variation \label{sec:spectralindex}}

As we noted in Section \ref{sec:intro}, the synchrotron spectral indices of NTFs vary significantly from filament to filament, have been measured along filaments, and tend to be steeper for higher latitude filaments. Since we propagate our \textmra{CRs} as a fluid, there is no way to extract spectral index values from the simulations directly. Here, we demonstrate a few antipodal effects that may qualitatively account for the variability.

Observations of elemental CR abundances show that the ratio of \textmra{CR} secondaries changes with their energy, which is usually interpreted as evidence for an energy dependent confinement time \citep{Evoli2020a}. If we model this dependence as a diffusivity which scales as a power law in energy, $\kappa(\gamma)\propto \gamma^q$, then the solution of the 1D spatial diffusion equation for impulsive injection is
\begin{equation}
    f(x,\gamma,t) = \frac{C\gamma^{-p}}{\sqrt{4\pi\kappa(\gamma)t}}\exp{\left( -\frac{x^2}{4\kappa(\gamma)t} \right)}\,.
\end{equation}

At a time $t$, each point $x$ along this 1-dimensional tube will have a different population spectral index, $p(\gamma)=-d\ln{f}/d\ln{\gamma}$. At each energy, the population has a "tangent power law." By defining a diffusion length scale $x_0(\gamma,t) = \sqrt{4\kappa(\gamma)t}$, we find the spectral index at a point $x$, at a time $t$, and at a frequency $\nu_c(\gamma)$ to be
\begin{equation}
    -\frac{d\ln{f}}{d\ln{\gamma}} = p + \frac{q}{2} - \frac{x^2}{4\kappa(\gamma)t}q = p + \frac{q}{2} - q\frac{x^2}{x_0(\nu_c(\gamma),t)^2}\,.
\end{equation}

\begin{figure}
    \centering
    \includegraphics[width=0.9\linewidth]{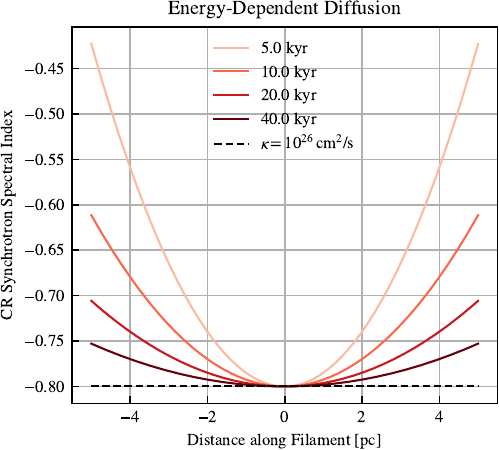}
    \caption{Modeled spectral index due to energy-dependent diffusion at different ages with $\kappa=10^{26}\cmcms$ at this energy band. Here, we set the original power law spectrum to be $p=2.5$ and the energy-dependent diffusion index to be $q=0.2$ in order to get an equilibrium population index that matches 2.6,  the power law index we assumed without energy-dependent diffusion.}
    \label{fig:E-diffusion}
\end{figure}

Assuming that the primary emission is synchrotron radiation, and that each \textmra{CR} emits at its synchrotron critical frequency, then the synchrotron spectral index is $s(\gamma) = (p(\gamma)-1)/2$. We therefore see that the synchrotron spectral index we observe depends not only on the frequency band and age of the filaments, but also on displacement along the filament. The index at earlier ages near the outskirts of the filament will be much flatter than the center value, because the population far from the injection site will be skewed towards the more energetic particles. Evolution of the synchrotron spectral index along a filament is depicted in Figure \ref{fig:E-diffusion}, where a general tread of spectral hardening along the filament and a harder mean spectrum seems clear. Observations of synchrotron spectral index could be fitted to these parabolas to constrain the ages and diffusivity in the filaments. We note that although we have not modeled it, a similar trend would be observed in a streaming dominated filament, due to the weakening of self confinement and larger streaming velocity expected at higher energy \citep{Zweibel2017}.

It is well known that synchrotron radiation also provides a method of altering the spectral index, though in the direction opposite to energy dependent diffusion. The energy lost by higher CR energies is greater than the energy lost by lower CR energies, thus leading to a steepening of the spectrum over time. We quantify this by noting that synchrotron energy loss scales as $d\gamma/dt = 4\sigma_T\gamma^2 U_B/(3mc) = A\gamma^2$. Integrating this differential equation, we get the relations between a particle's time-dependent energy $\gamma(t)$ and its original energy $\gamma_0$ as
\begin{equation}\label{eq:gammagamma0}
    \gamma(t) = \frac{\gamma_0}{1+\gamma_0At}\,, \quad \frac{d\gamma}{d\gamma_0} = \frac{\gamma^2}{\gamma_0^2}\,.
\end{equation}

Any power law injection initially of the form $f_0(\gamma_0) = C\gamma_0^{-\alpha}$ will evolve as $f(\gamma)d\gamma = f_0(\gamma_0)d\gamma_0$. This, combined with Equation (\ref{eq:gammagamma0}), tells us that the resulting spectrum is
\begin{equation}\label{eq:population_time}
    f(\gamma,t) = C\gamma_0^{-\alpha}\frac{\gamma_0^2}{\gamma^2} = C\gamma^{-\alpha}(1-\alpha At)^{\alpha-2}\,.
\end{equation}

\begin{figure}
    \centering
    \includegraphics[width=\linewidth]{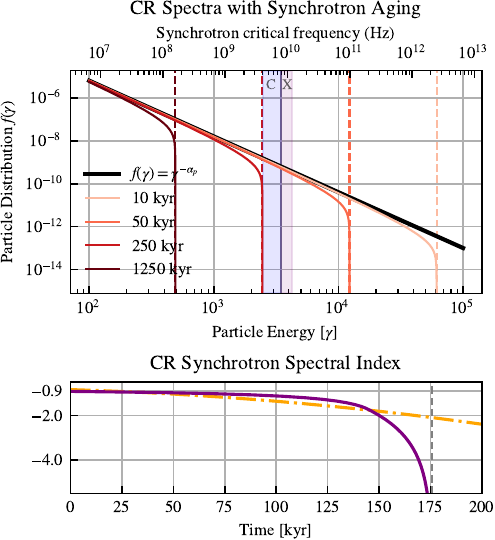}
    \caption{Plotted are \textmra{CR} distribution functions as a function of energy at different ages. Only synchrotron losses are assumed in these predictions. In the bottom plot, the CR synchrotron spectral index is plotted as a function of time between the VLA C-- and X--bands that cover 4--8 GHz (blue region) and 8--12 GHz (pink region) respectively. The vertical asymptote shown is the time when there are no more particles left that emit in the X--band. The dashdotted line shows a numerical calculation of the spectral index using the complete Bessel function treatment, showing that the spectral index would continue decreasing beyond the asymptote.}
    \label{fig:spectra_CR}
\end{figure}

Snapshots of the CRe$^\pm$ population spectrum given in Equation (\ref{eq:population_time}) are plotted in the top graph of Figure \ref{fig:spectra_CR}, given a magnetic field strength of $200\muG$. In the bottom graph, we show the spectral index that would be derived from observing emission with the C-- and X-- VLA bands that cover 4--8 GHz and 8--12 GHz respectively. This is quantified by fitting the emission per frequency $P_\nu$ to a power law with $\gamma$. We accomplish this by knowing that $P_\nu\propto \gamma^2f(\gamma)$, so we obtain the synchrotron spectral index as
\begin{align}\label{eq:spectralindex}
    s(t) = \frac{d\ln P_\nu(t)}{d \mathmra{\,\ln\,} \nu}\approx \frac{\ln\left(\frac{\int_{\gamma(8 \GHz)}^{\gamma(12\GHz)}\gamma^2f(\gamma,t)d\gamma}{\int_{\gamma(4 \GHz)}^{\gamma(8\GHz)}\gamma^2f(\gamma,t)d\gamma}\right)}{\ln\left(\frac{\int_{8 \GHz}^{12 \GHz}d\nu}{\int_{4 \GHz}^{8 \GHz}d\nu}\right)}\,.
\end{align}

We find that through this effect, the synchrotron spectral index will be strictly decreasing over time, and at 175 kyr, the emission will vanish due to the asymptoting \textmra{CR} population. We ran a numerical simulation using the complete treatment of synchrotron radiation \citep{RL} and found a similar result, though the vertical asymptote  vanished. This was because particles were  allowed to emit outside of their critical frequency. The emission is this regime was greatly decreased, however, so our conclusions remain the same.

Finally, secondary production can be yet another cause of spectral steepening, as the source function is generally steeper than the injection spectrum \citep{DolagEnsslin}. Specifically, for an initial proton population index of -2.6, the steady-state electron spectrum will have a population index of -3.8. This steady-state population will dominate at timescales longer than $\tau_{tot}$ and $\tau_{sec}$ from Equations (\ref{eq:loss2})\&(\ref{eq:secondary_production}).

The relative importance of these effects depends on the age of the filament and the properties of the background medium as well as the \textmra{CR} source itself. For our fiducial parameters, synchrotron aging and secondary production come into play at much later times than energy dependent propagation or secondary production (which of course requires injection of \textmra{CR} protons). However, in a denser filament with a stronger magnetic field, this ordering could be reversed.

Yet, as mentioned before, NTF observations show a very broad distribution of their synchrotron spectral indices. Not only do their indices range from -2 to +0.5, but along a filament, the index appears jagged without any clear symmetry point \citep{Law2008, Pare2022}. For many of these NTFs, there is no combination of effects we have presented that can fully explain the observations. It is clear that there are factors---either propagation-related or source-driven---yet to be accounted for to completely describe an NTF's spectral index.

\subsection{Turbulent Origin Mechanism}

\citet{Yusef2022c} \citet{Yusef2022a} find that, on average, approximately one stellar IR source should exist in proximity to any specific filament that is the length and width of ours. This, alongside observational evidence of compact sources interacting with specific NTFs, such as the Wishbone, gives credence to the theory of CR injection via compact sources \citep{Zhao2025}. At the same time, there exist certain filaments that are not proximate to any such compact source, so this mechanism cannot be the end of the story \citep{Pare2022}. It becomes prudent to examine other origin models that are unconstrained by the requirement of a source object within causal range of the NTF.

Throughout the simulation, we have been assuming a perfectly straight flux tube with no magnetic field variations. This justified the 1D approach without needing to vary the width of the flux tube along the filament. The true situation is more complicated: various instabilities within the plasma and the Milky Way will act to disturb order in the disk, and it is highly likely that a flux tube will not be purely straight nor have a uniform magnetic field. Many observed filaments do indeed have these imperfections, showing as kinks, splits, and multiple peaks in brightness over their length. In addition to their orientations not being perfectly perpendicular to the Galactic Plane, these factors indicate a complex magnetic field structure within the Galactic Center \citep{Yusef2022b, Nord2004}.

A slight perturbation in the topology of the magnetic field could result in various propagation effects that cause CRs to deviate from the 1D diffusion equation. For instance, if the magnetic field lines comprising the filament were to fray outward at some distance away from the ``origin,'' this would reduce the magnetic field strength by the flux freezing law, thus changing the critical $\gamma$ that the synchrotron emission tracks, thus reducing the emissivity. These effects could also explain a particularly plateaued filament, as a completely uniform \textmra{CR} energy density can exhibit signs of ``propagating'' by simply varying the radius of the NTF along its length.

\citet{Boldyrev} proposed a turbulent origin for the NTFs. Compared to the surrounding magnetic field, the NTF structures appear to be amplified in strength by two orders of magnitude, and their axis is often slightly misaligned with a poloidal magnetic field. These effects can arise from inhomogeneous MHD turbulence, as regions of stronger turbulence can create and expel large amounts of magnetic flux \citep{Zhou2024}. The turbulent strain rate is capable of stretching the magnetic field eddies into thin filaments when the magnetic diffusivity is small \citep{TennekesLumley}. In the case of these NTFs, they could have been created from various turbulent drivers in the Galactic Plane that then create a magnetic dynamo effect and expel the filaments as byproducts \citep{VlahosIsliker}. \textmra{These dynamical processes could also create a multi-phase ISM, which would allow NTFs to exist at temperatures and densities that are not representative of the surrounding CMZ \citep{Ng2026}.} The addition of \textmra{CRs} contributes a significant amount of feedback on ISM turbulence as well, leading to more observables \citep{Habegger2024}. We intend to pursue future work to simulate these turbulent effects and explore this intriguing hypothesis for the origin of NTFs.

\section{Conclusions \label{sec:conclusion}}

This study aimed to find observable differences between injection cases including and excluding protons from the CR population. While a large but radiatively invisible injection of \textmra{CR} protons drives a stronger flow along the filament and provides more heating than the observed population of \textmra{CR} electrons, these effects are too small to be observable, even at the lowest background densities we considered (see Figures \ref{fig:EcLc} \& \ref{fig:vel1heating_p}). The most consequential effects of protons may be that they makes the self confinement model more viable (see Appendix \ref{sec:selfconfinement}), and, at background densities of several hundred $\cm^{-3}$ or more, they provide secondary lepton \textmra{CRs} through hadronic interactions. Our simulations have shown that the energy profile of each appear to be a strict scaling of each other by the proton-to-electron ratio, and the luminosity profiles being mostly identical. 

Our modeling did reveal a number of interesting effects. We have found that the shape of the energy profiles are closely tied to the diffusion and streaming parameters chosen for each simulation, and that it may be possible to extract values for $\kappa$ and $v_A$ from observations.

Our calculations have shown that the length of a filament is not limited by radiative or collisional losses: as a filament grows older, the filament will grow progressively longer. The absence of filaments longer than a few 10s of pc suggests that the length may be limited by the underlying magnetic topology, suggesting that attempts to explain the filaments as manifestations of MHD turbulence \citep{Boldyrev} warrant further investigation. \textmra{This anomalous NTF length may also suggest that alternate CR propagation methods, such as spatially-dependent diffusion, are at play within the CMZ \citep{Ewart2026}}

This time-dependent length still provides valuable information, as mentioned above through $\kappa$ and $v_A$ constraints, thus allowing us to obtain the age of a filament if we know the former two parameters well. Another route to solving for the age of a filament is through observations of its spectral indices, where we present two methods in which the synchrotron spectral index will change over time. Fitting observations to these two methods could result in more data to constrain the age and origin of a particular filament.

Our future work will investigate turbulence as an originator for filaments. \cite{Boldyrev} have previously shown that the observations of the NTFs and their surroundings imply a chaotic, locally amplified magnetic field as opposed to a uniform poloidal magnetic field. This claim is recently backed up by \cite{Zhao2025}, who have observed that the magnetic field has a background $\sim30\muG$ magnetic field, and the NTFs lie on local amplifications of $\sim1\mG$ field lines. Our planned work involves 3D simulations of the turbulent ISM with \textmra{CR} propagation and acceleration to test this origin theory of NTFs.

\begin{acknowledgments}
This project was funded by NASA award \#09-0054, NSF awards 2007323 and 240922, the Simons Collaboration for Extreme Electrodynamics of Compact Sources, and the University of Wisconsin-Madison, including the Community of Graduate Research Scholars Fellowship. \textmra{Figure \ref{fig:LIC} uses data from the MeerKAT telescope operated by the South African Radio Astronomy Observatory, which is a facility of the National Research Foundation, an agency of the Department of Science and Innovation.} We would like to thank the FIREPLACE collaboration for their support and input. \textmra{We also thank the anonymous referee for their insightful questions and comments that furthered the breadth of the work and its implications.} E.Z. is grateful for the hospitality of the Leonard Parker Center for Cosmology and Gravitational Astrophysics at U. Wisconsin-Milwaukee, where a portion of this work was carried out.
\end{acknowledgments}

\software{\texttt{Athena++} \citep{Stone2020, JiangOh}, \texttt{MatPlotLib} \citep{Hunter2007}, \texttt{NumPy} \citep{vanderWalt2011,Harris2020}, \texttt{AstroPy} \citep{Astropy2013,Astropy2018}}

\bibliography{sample7}{}

@ARTICLE{SM,
       author = {{Serabyn}, E. and {Morris}, Mark},
        title = "{The Source of the Relativistic Particles in the Galactic Center Arc}",
      journal = {\apjl},
         year = 1994,
        month = apr,
       volume = {424},
        pages = {L91},
          doi = {10.1086/187282},
       adsurl = {https://ui.adsabs.harvard.edu/abs/1994ApJ...424L..91S}
}

@ARTICLE{SchlickeiserMannheim,       author = {{Mannheim}, K. and {Schlickeiser}, R.},
        title = "{Interactions of cosmic ray nuclei}",
      journal = {\aap},
         year = 1994,
        month = jun,
       volume = {286},
        pages = {983-996},
          doi = {10.48550/arXiv.astro-ph/9402042},
    archivePrefix = {arXiv},
       eprint = {astro-ph/9402042},
    primaryClass = {astro-ph},
       adsurl = {https://ui.adsabs.harvard.edu/abs/1994A&A...286..983M}
}

@BOOK{RL,
       author = {{Rybicki}, George B. and {Lightman}, Alan P.},
    publisher = {Wiley-Interscience},
        title = "{Radiative Processes in Astrophysics}",
         year = 1986,
       adsurl = {https://ui.adsabs.harvard.edu/abs/1986rpa..book.....R}
}

@ARTICLE{DolagEnsslin,
       author = {{Dolag}, K. and {En{\ss}lin}, T.~A.},
        title = "{Radio halos of galaxy clusters from hadronic secondary electron injection in realistic magnetic field configurations}",
      journal = {\aap},
         year = 2000,
        month = oct,
       volume = {362},
        pages = {151-157},
          doi = {10.48550/arXiv.astro-ph/0008333},
    archivePrefix = {arXiv},
       eprint = {astro-ph/0008333},
    primaryClass = {astro-ph},
       adsurl = {https://ui.adsabs.harvard.edu/abs/2000A&A...362..151D}
}

@ARTICLE{TPE,
       author = {{Thomas}, Timon and {Pfrommer}, Christoph and {En{\ss}lin}, Torsten},
        title = "{Probing Cosmic-Ray Transport with Radio Synchrotron Harps in the Galactic Center}",
      journal = {\apjl},
         year = 2020,
        month = feb,
       volume = {890},
       number = {2},
          eid = {L18},
        pages = {L18},
          doi = {10.3847/2041-8213/ab7237},
    archivePrefix = {arXiv},
       eprint = {1912.08491},
    primaryClass = {astro-ph.HE},
       adsurl = {https://ui.adsabs.harvard.edu/abs/2020ApJ...890L..18T}
}

@ARTICLE{JiangOh,
       author = {{Jiang}, Yan-Fei and {Oh}, S. Peng},
        title = "{A New Numerical Scheme for Cosmic-Ray Transport}",
      journal = {\apj},
         year = 2018,
        month = feb,
       volume = {854},
       number = {1},
          eid = {5},
        pages = {5},
          doi = {10.3847/1538-4357/aaa6ce},
archivePrefix = {arXiv},
       eprint = {1712.07117},
 primaryClass = {astro-ph.HE},
       adsurl = {https://ui.adsabs.harvard.edu/abs/2018ApJ...854....5J}
}

@ARTICLE{YH,
       author = {{Yoast-Hull}, Tova M. and {Gallagher}, III, J.~S. and {Zweibel}, Ellen G.},
        title = "{The Cosmic-Ray Population of the Galactic Central Molecular Zone}",
      journal = {\apj},
         year = 2014,
        month = aug,
       volume = {790},
       number = {2},
          eid = {86},
        pages = {86},
          doi = {10.1088/0004-637X/790/2/86},
archivePrefix = {arXiv},
       eprint = {1405.7059},
 primaryClass = {astro-ph.HE},
       adsurl = {https://ui.adsabs.harvard.edu/abs/2014ApJ...790...86Y}
}

@ARTICLE{Pare2022,
       author = {{Par{\'e}}, Dylan M. and {Lang}, Cornelia C. and {Morris}, Mark R.},
        title = "{A Very Large Array Study of Newly Discovered Southern Latitude Nonthermal Filaments in the Galactic Center: Radio Continuum Total-intensity and Spectral Index Properties}",
      journal = {\apj},
         year = 2022,
        month = dec,
       volume = {941},
       number = {2},
          eid = {123},
        pages = {123},
          doi = {10.3847/1538-4357/aca40a},
archivePrefix = {arXiv},
       eprint = {2209.08153},
 primaryClass = {astro-ph.GA},
       adsurl = {https://ui.adsabs.harvard.edu/abs/2022ApJ...941..123P}
}

@BOOK{Schlickeiser2002,
       author = {{Schlickeiser}, Reinhard},
        title = "{Cosmic Ray Astrophysics}",
         year = 2002,
       adsurl = {https://ui.adsabs.harvard.edu/abs/2002cra..book.....S},
    publisher = {Springer}
}

@ARTICLE{Stone2020,
       author = {{Stone}, James M. and {Tomida}, Kengo and {White}, Christopher J. and {Felker}, Kyle G.},
        title = "{The Athena++ Adaptive Mesh Refinement Framework: Design and Magnetohydrodynamic Solvers}",
      journal = {\apjs},
         year = 2020,
        month = jul,
       volume = {249},
       number = {1},
          eid = {4},
        pages = {4},
          doi = {10.3847/1538-4365/ab929b},
archivePrefix = {arXiv},
       eprint = {2005.06651},
 primaryClass = {astro-ph.IM},
       adsurl = {https://ui.adsabs.harvard.edu/abs/2020ApJS..249....4S}
}

@ARTICLE{Hunter2007,
       author = {{Hunter}, John D.},
        title = "{Matplotlib: A 2D Graphics Environment}",
      journal = {Computing in Science and Engineering},
         year = 2007,
        month = may,
       volume = {9},
       number = {3},
        pages = {90-95},
          doi = {10.1109/MCSE.2007.55},
       adsurl = {https://ui.adsabs.harvard.edu/abs/2007CSE.....9...90H}
}

@ARTICLE{vanderWalt2011,
       author = {{van der Walt}, St{\'e}fan and {Colbert}, S. Chris and {Varoquaux}, Ga{\"e}l},
        title = "{The NumPy Array: A Structure for Efficient Numerical Computation}",
      journal = {Computing in Science and Engineering},
         year = 2011,
        month = mar,
       volume = {13},
       number = {2},
        pages = {22-30},
          doi = {10.1109/MCSE.2011.37},
archivePrefix = {arXiv},
       eprint = {1102.1523},
 primaryClass = {cs.MS},
       adsurl = {https://ui.adsabs.harvard.edu/abs/2011CSE....13b..22V}
}

@ARTICLE{Harris2020,
       author = {{Harris}, Charles R. and {Millman}, K. Jarrod and {van der Walt}, St{\'e}fan J. and {Gommers}, Ralf and {Virtanen}, Pauli and {Cournapeau}, David and {Wieser}, Eric and {Taylor}, Julian and {Berg}, Sebastian and {Smith}, Nathaniel J. and {Kern}, Robert and {Picus}, Matti and {Hoyer}, Stephan and {van Kerkwijk}, Marten H. and {Brett}, Matthew and {Haldane}, Allan and {del R{\'\i}o}, Jaime Fern{\'a}ndez and {Wiebe}, Mark and {Peterson}, Pearu and {G{\'e}rard-Marchant}, Pierre and {Sheppard}, Kevin and {Reddy}, Tyler and {Weckesser}, Warren and {Abbasi}, Hameer and {Gohlke}, Christoph and {Oliphant}, Travis E.},
        title = "{Array programming with NumPy}",
      journal = {\nat},
         year = 2020,
        month = sep,
       volume = {585},
       number = {7825},
        pages = {357-362},
          doi = {10.1038/s41586-020-2649-2},
archivePrefix = {arXiv},
       eprint = {2006.10256},
 primaryClass = {cs.MS},
       adsurl = {https://ui.adsabs.harvard.edu/abs/2020Natur.585..357H}
}

@ARTICLE{Astropy2013,
       author = {{Astropy Collaboration} and {Robitaille}, Thomas P. and {Tollerud}, Erik J. and {Greenfield}, Perry and {Droettboom}, Michael and {Bray}, Erik and {Aldcroft}, Tom and {Davis}, Matt and {Ginsburg}, Adam and {Price-Whelan}, Adrian M. and {Kerzendorf}, Wolfgang E. and {Conley}, Alexander and {Crighton}, Neil and {Barbary}, Kyle and {Muna}, Demitri and {Ferguson}, Henry and {Grollier}, Fr{\'e}d{\'e}ric and {Parikh}, Madhura M. and {Nair}, Prasanth H. and {Unther}, Hans M. and {Deil}, Christoph and {Woillez}, Julien and {Conseil}, Simon and {Kramer}, Roban and {Turner}, James E.~H. and {Singer}, Leo and {Fox}, Ryan and {Weaver}, Benjamin A. and {Zabalza}, Victor and {Edwards}, Zachary I. and {Azalee Bostroem}, K. and {Burke}, D.~J. and {Casey}, Andrew R. and {Crawford}, Steven M. and {Dencheva}, Nadia and {Ely}, Justin and {Jenness}, Tim and {Labrie}, Kathleen and {Lim}, Pey Lian and {Pierfederici}, Francesco and {Pontzen}, Andrew and {Ptak}, Andy and {Refsdal}, Brian and {Servillat}, Mathieu and {Streicher}, Ole},
        title = "{Astropy: A community Python package for astronomy}",
      journal = {\aap},
         year = 2013,
        month = oct,
       volume = {558},
          eid = {A33},
        pages = {A33},
          doi = {10.1051/0004-6361/201322068},
archivePrefix = {arXiv},
       eprint = {1307.6212},
 primaryClass = {astro-ph.IM},
       adsurl = {https://ui.adsabs.harvard.edu/abs/2013A&A...558A..33A}
}

@ARTICLE{Astropy2018,
       author = {{Astropy Collaboration} and {Price-Whelan}, A.~M. and {Sip{\H{o}}cz}, B.~M. and {G{\"u}nther}, H.~M. and {Lim}, P.~L. and {Crawford}, S.~M. and {Conseil}, S. and {Shupe}, D.~L. and {Craig}, M.~W. and {Dencheva}, N. and {Ginsburg}, A. and {VanderPlas}, J.~T. and {Bradley}, L.~D. and {P{\'e}rez-Su{\'a}rez}, D. and {de Val-Borro}, M. and {Aldcroft}, T.~L. and {Cruz}, K.~L. and {Robitaille}, T.~P. and {Tollerud}, E.~J. and {Ardelean}, C. and {Babej}, T. and {Bach}, Y.~P. and {Bachetti}, M. and {Bakanov}, A.~V. and {Bamford}, S.~P. and {Barentsen}, G. and {Barmby}, P. and {Baumbach}, A. and {Berry}, K.~L. and {Biscani}, F. and {Boquien}, M. and {Bostroem}, K.~A. and {Bouma}, L.~G. and {Brammer}, G.~B. and {Bray}, E.~M. and {Breytenbach}, H. and {Buddelmeijer}, H. and {Burke}, D.~J. and {Calderone}, G. and {Cano Rodr{\'\i}guez}, J.~L. and {Cara}, M. and {Cardoso}, J.~V.~M. and {Cheedella}, S. and {Copin}, Y. and {Corrales}, L. and {Crichton}, D. and {D'Avella}, D. and {Deil}, C. and {Depagne}, {\'E}. and {Dietrich}, J.~P. and {Donath}, A. and {Droettboom}, M. and {Earl}, N. and {Erben}, T. and {Fabbro}, S. and {Ferreira}, L.~A. and {Finethy}, T. and {Fox}, R.~T. and {Garrison}, L.~H. and {Gibbons}, S.~L.~J. and {Goldstein}, D.~A. and {Gommers}, R. and {Greco}, J.~P. and {Greenfield}, P. and {Groener}, A.~M. and {Grollier}, F. and {Hagen}, A. and {Hirst}, P. and {Homeier}, D. and {Horton}, A.~J. and {Hosseinzadeh}, G. and {Hu}, L. and {Hunkeler}, J.~S. and {Ivezi{\'c}}, {\v{Z}}. and {Jain}, A. and {Jenness}, T. and {Kanarek}, G. and {Kendrew}, S. and {Kern}, N.~S. and {Kerzendorf}, W.~E. and {Khvalko}, A. and {King}, J. and {Kirkby}, D. and {Kulkarni}, A.~M. and {Kumar}, A. and {Lee}, A. and {Lenz}, D. and {Littlefair}, S.~P. and {Ma}, Z. and {Macleod}, D.~M. and {Mastropietro}, M. and {McCully}, C. and {Montagnac}, S. and {Morris}, B.~M. and {Mueller}, M. and {Mumford}, S.~J. and {Muna}, D. and {Murphy}, N.~A. and {Nelson}, S. and {Nguyen}, G.~H. and {Ninan}, J.~P. and {N{\"o}the}, M. and {Ogaz}, S. and {Oh}, S. and {Parejko}, J.~K. and {Parley}, N. and {Pascual}, S. and {Patil}, R. and {Patil}, A.~A. and {Plunkett}, A.~L. and {Prochaska}, J.~X. and {Rastogi}, T. and {Reddy Janga}, V. and {Sabater}, J. and {Sakurikar}, P. and {Seifert}, M. and {Sherbert}, L.~E. and {Sherwood-Taylor}, H. and {Shih}, A.~Y. and {Sick}, J. and {Silbiger}, M.~T. and {Singanamalla}, S. and {Singer}, L.~P. and {Sladen}, P.~H. and {Sooley}, K.~A. and {Sornarajah}, S. and {Streicher}, O. and {Teuben}, P. and {Thomas}, S.~W. and {Tremblay}, G.~R. and {Turner}, J.~E.~H. and {Terr{\'o}n}, V. and {van Kerkwijk}, M.~H. and {de la Vega}, A. and {Watkins}, L.~L. and {Weaver}, B.~A. and {Whitmore}, J.~B. and {Woillez}, J. and {Zabalza}, V. and {Astropy Contributors}},
        title = "{The Astropy Project: Building an Open-science Project and Status of the v2.0 Core Package}",
      journal = {\aj},
         year = 2018,
        month = sep,
       volume = {156},
       number = {3},
          eid = {123},
        pages = {123},
          doi = {10.3847/1538-3881/aabc4f},
archivePrefix = {arXiv},
       eprint = {1801.02634},
 primaryClass = {astro-ph.IM},
       adsurl = {https://ui.adsabs.harvard.edu/abs/2018AJ....156..123A}
}

@BOOK{DM,
       author = {{Dermer}, Charles D. and {Menon}, Govind},
        title = "{High Energy Radiation from Black Holes: Gamma Rays, Cosmic Rays, and Neutrinos}",
         year = 2009,
       adsurl = {https://ui.adsabs.harvard.edu/abs/2009herb.book.....D},
    publisher = {Princeton University Press}
}

@ARTICLE{Law2008,
       author = {{Law}, C.~J. and {Yusef-Zadeh}, F. and {Cotton}, W.~D.},
        title = "{A Wide-Area VLA Continuum Survey near the Galactic Center at 6 and 20 cm Wavelengths}",
      journal = {\apjs},
         year = 2008,
        month = aug,
       volume = {177},
       number = {2},
        pages = {515-545},
          doi = {10.1086/588218},
archivePrefix = {arXiv},
       eprint = {0803.1412},
 primaryClass = {astro-ph},
       adsurl = {https://ui.adsabs.harvard.edu/abs/2008ApJS..177..515L}
}

@ARTICLE{Yusef2022a,
       author = {{Yusef-Zadeh}, F. and {Arendt}, R.~G. and {Wardle}, M. and {Heywood}, I. and {Cotton}, W. and {Camilo}, F.},
        title = "{Statistical Properties of the Population of the Galactic Center Filaments: the Spectral Index and Equipartition Magnetic Field}",
      journal = {\apjl},
         year = 2022,
        month = feb,
       volume = {925},
       number = {2},
          eid = {L18},
        pages = {L18},
          doi = {10.3847/2041-8213/ac4802},
archivePrefix = {arXiv},
       eprint = {2201.10552},
 primaryClass = {astro-ph.GA},
       adsurl = {https://ui.adsabs.harvard.edu/abs/2022ApJ...925L..18Y}
}

@ARTICLE{Boldyrev,
       author = {{Boldyrev}, Stanislav and {Yusef-Zadeh}, Farhad},
        title = "{Turbulent Origin of the Galactic Center Magnetic Field: Nonthermal Radio Filaments}",
      journal = {\apjl},
         year = 2006,
        month = feb,
       volume = {637},
       number = {2},
        pages = {L101-L104},
          doi = {10.1086/500411},
archivePrefix = {arXiv},
       eprint = {astro-ph/0512373},
 primaryClass = {astro-ph},
       adsurl = {https://ui.adsabs.harvard.edu/abs/2006ApJ...637L.101B}
}

@ARTICLE{YZ1984,
       author = {{Yusef-Zadeh}, F. and {Morris}, M. and {Chance}, D.},
        title = "{Large, highly organized radio structures near the galactic centre}",
      journal = {\nat},
         year = 1984,
        month = aug,
       volume = {310},
       number = {5978},
        pages = {557-561},
          doi = {10.1038/310557a0},
       adsurl = {https://ui.adsabs.harvard.edu/abs/1984Natur.310..557Y}
}

@ARTICLE{Heywood2022,
       author = {{Heywood}, I. and {Rammala}, I. and {Camilo}, F. and {Cotton}, W.~D. and {Yusef-Zadeh}, F. and {Abbott}, T.~D. and {Adam}, R.~M. and {Adams}, G. and {Aldera}, M.~A. and {Asad}, K.~M.~B. and {Bauermeister}, E.~F. and {Bennett}, T.~G.~H. and {Bester}, H.~L. and {Bode}, W.~A. and {Botha}, D.~H. and {Botha}, A.~G. and {Brederode}, L.~R.~S. and {Buchner}, S. and {Burger}, J.~P. and {Cheetham}, T. and {de Villiers}, D.~I.~L. and {Dikgale-Mahlakoana}, M.~A. and {du Toit}, L.~J. and {Esterhuyse}, S.~W.~P. and {Fanaroff}, B.~L. and {February}, S. and {Fourie}, D.~J. and {Frank}, B.~S. and {Gamatham}, R.~R.~G. and {Geyer}, M. and {Goedhart}, S. and {Gouws}, M. and {Gumede}, S.~C. and {Hlakola}, M.~J. and {Hokwana}, A. and {Hoosen}, S.~W. and {Horrell}, J.~M.~G. and {Hugo}, B. and {Isaacson}, A.~I. and {J{\'o}zsa}, G.~I.~G. and {Jonas}, J.~L. and {Joubert}, A.~F. and {Julie}, R.~P.~M. and {Kapp}, F.~B. and {Kenyon}, J.~S. and {Kotz{\'e}}, P.~P.~A. and {Kriek}, N. and {Kriel}, H. and {Krishnan}, V.~K. and {Lehmensiek}, R. and {Liebenberg}, D. and {Lord}, R.~T. and {Lunsky}, B.~M. and {Madisa}, K. and {Magnus}, L.~G. and {Mahgoub}, O. and {Makhaba}, A. and {Makhathini}, S. and {Malan}, J.~A. and {Manley}, J.~R. and {Marais}, S.~J. and {Martens}, A. and {Mauch}, T. and {Merry}, B.~C. and {Millenaar}, R.~P. and {Mnyandu}, N. and {Mokone}, O.~J. and {Monama}, T.~E. and {Mphego}, M.~C. and {New}, W.~S. and {Ngcebetsha}, B. and {Ngoasheng}, K.~J. and {Ockards}, M.~T. and {Oozeer}, N. and {Otto}, A.~J. and {Passmoor}, S.~S. and {Patel}, A.~A. and {Peens-Hough}, A. and {Perkins}, S.~J. and {Ramaila}, A.~J.~T. and {Ramanujam}, N.~M.~R. and {Ramudzuli}, Z.~R. and {Ratcliffe}, S.~M. and {Robyntjies}, A. and {Salie}, S. and {Sambu}, N. and {Schollar}, C.~T.~G. and {Schwardt}, L.~C. and {Schwartz}, R.~L. and {Serylak}, M. and {Siebrits}, R. and {Sirothia}, S.~K. and {Slabber}, M. and {Smirnov}, O.~M. and {Sofeya}, L. and {Taljaard}, B. and {Tasse}, C. and {Tiplady}, A.~J. and {Toruvanda}, O. and {Twum}, S.~N. and {van Balla}, T.~J. and {van der Byl}, A. and {van der Merwe}, C. and {Van Tonder}, V. and {Van Wyk}, R. and {Venter}, A.~J. and {Venter}, M. and {Wallace}, B.~H. and {Welz}, M.~G. and {Williams}, L.~P. and {Xaia}, B.},
        title = "{The 1.28 GHz MeerKAT Galactic Center Mosaic}",
      journal = {\apj},
         year = 2022,
        month = feb,
       volume = {925},
       number = {2},
          eid = {165},
        pages = {165},
          doi = {10.3847/1538-4357/ac449a},
archivePrefix = {arXiv},
       eprint = {2201.10541},
 primaryClass = {astro-ph.GA},
       adsurl = {https://ui.adsabs.harvard.edu/abs/2022ApJ...925..165H}
}

@ARTICLE{Habegger2024,
       author = {{Habegger}, Roark and {Ho}, Ka Wai and {Yuen}, Ka Ho and {Zweibel}, Ellen G.},
        title = "{Cosmic-Ray Feedback on Bistable Interstellar Medium Turbulence}",
      journal = {\apj},
         year = 2024,
        month = oct,
       volume = {974},
       number = {1},
          eid = {17},
        pages = {17},
          doi = {10.3847/1538-4357/ad67da},
archivePrefix = {arXiv},
       eprint = {2403.07976},
 primaryClass = {astro-ph.HE},
       adsurl = {https://ui.adsabs.harvard.edu/abs/2024ApJ...974...17H}
}

@ARTICLE{Morris1985,
       author = {{Morris}, M. and {Yusef-Zadeh}, F.},
        title = "{Unusual threads of radio emission near the Galactic Center.}",
      journal = {\aj},
         year = 1985,
        month = dec,
       volume = {90},
        pages = {2511-2513},
          doi = {10.1086/113955},
       adsurl = {https://ui.adsabs.harvard.edu/abs/1985AJ.....90.2511M}
}

@ARTICLE{Bally1989,
       author = {{Bally}, John and {Yusef-Zadeh}, Farhad},
        title = "{A New System of Nonthermal Filaments near the Galactic Center}",
      journal = {\apj},
         year = 1989,
        month = jan,
       volume = {336},
        pages = {173},
          doi = {10.1086/167003},
       adsurl = {https://ui.adsabs.harvard.edu/abs/1989ApJ...336..173B}
}

@ARTICLE{Heywood2019,
       author = {{Heywood}, I. and {Camilo}, F. and {Cotton}, W.~D. and {Yusef-Zadeh}, F. and {Abbott}, T.~D. and {Adam}, R.~M. and {Aldera}, M.~A. and {Bauermeister}, E.~F. and {Booth}, R.~S. and {Botha}, A.~G. and {Botha}, D.~H. and {Brederode}, L.~R.~S. and {Brits}, Z.~B. and {Buchner}, S.~J. and {Burger}, J.~P. and {Chalmers}, J.~M. and {Cheetham}, T. and {de Villiers}, D. and {Dikgale-Mahlakoana}, M.~A. and {du Toit}, L.~J. and {Esterhuyse}, S.~W.~P. and {Fanaroff}, B.~L. and {Foley}, A.~R. and {Fourie}, D.~J. and {Gamatham}, R.~R.~G. and {Goedhart}, S. and {Gounden}, S. and {Hlakola}, M.~J. and {Hoek}, C.~J. and {Hokwana}, A. and {Horn}, D.~M. and {Horrell}, J.~M.~G. and {Hugo}, B. and {Isaacson}, A.~R. and {Jonas}, J.~L. and {Jordaan}, J.~D.~B.~L. and {Joubert}, A.~F. and {J{\'o}zsa}, G.~I.~G. and {Julie}, R.~P.~M. and {Kapp}, F.~B. and {Kenyon}, J.~S. and {Kotz{\'e}}, P.~P.~A. and {Kriel}, H. and {Kusel}, T.~W. and {Lehmensiek}, R. and {Liebenberg}, D. and {Loots}, A. and {Lord}, R.~T. and {Lunsky}, B.~M. and {Macfarlane}, P.~S. and {Magnus}, L.~G. and {Magozore}, C.~M. and {Mahgoub}, O. and {Main}, J.~P.~L. and {Malan}, J.~A. and {Malgas}, R.~D. and {Manley}, J.~R. and {Maree}, M.~D.~J. and {Merry}, B. and {Millenaar}, R. and {Mnyandu}, N. and {Moeng}, I.~P.~T. and {Monama}, T.~E. and {Mphego}, M.~C. and {New}, W.~S. and {Ngcebetsha}, B. and {Oozeer}, N. and {Otto}, A.~J. and {Passmoor}, S.~S. and {Patel}, A.~A. and {Peens-Hough}, A. and {Perkins}, S.~J. and {Ratcliffe}, S.~M. and {Renil}, R. and {Rust}, A. and {Salie}, S. and {Schwardt}, L.~C. and {Serylak}, M. and {Siebrits}, R. and {Sirothia}, S.~K. and {Smirnov}, O.~M. and {Sofeya}, L. and {Swart}, P.~S. and {Tasse}, C. and {Taylor}, D.~T. and {Theron}, I.~P. and {Thorat}, K. and {Tiplady}, A.~J. and {Tshongweni}, S. and {van Balla}, T.~J. and {van der Byl}, A. and {van der Merwe}, C. and {van Dyk}, C.~L. and {Van Rooyen}, R. and {Van Tonder}, V. and {Van Wyk}, R. and {Wallace}, B.~H. and {Welz}, M.~G. and {Williams}, L.~P.},
        title = "{Inflation of 430-parsec bipolar radio bubbles in the Galactic Centre by an energetic event}",
      journal = {\nat},
         year = 2019,
        month = sep,
       volume = {573},
       number = {7773},
        pages = {235-237},
          doi = {10.1038/s41586-019-1532-5},
archivePrefix = {arXiv},
       eprint = {1909.05534},
 primaryClass = {astro-ph.GA},
       adsurl = {https://ui.adsabs.harvard.edu/abs/2019Natur.573..235H}
}

@BOOK{TennekesLumley,
       author = {{Tennekes}, H. and {Lumley}, J.~L.},
        title = "{First Course in Turbulence}",
    publisher = {MIT Press},
         year = 1972,
       adsurl = {https://ui.adsabs.harvard.edu/abs/1972fct..book.....T}
}

@ARTICLE{Zhou2024,
       author = {{Zhou}, Muni and {Zhdankin}, Vladimir and {Kunz}, Matthew W. and {Loureiro}, Nuno F. and {Uzdensky}, Dmitri A.},
        title = "{Magnetogenesis in a Collisionless Plasma: From Weibel Instability to Turbulent Dynamo}",
      journal = {\apj},
         year = 2024,
        month = jan,
       volume = {960},
       number = {1},
          eid = {12},
        pages = {12},
          doi = {10.3847/1538-4357/ad0b0f},
archivePrefix = {arXiv},
       eprint = {2308.01924},
 primaryClass = {astro-ph.HE},
       adsurl = {https://ui.adsabs.harvard.edu/abs/2024ApJ...960...12Z}
}

@ARTICLE{VlahosIsliker,
       author = {{Vlahos}, Loukas and {Isliker}, Heinz},
        title = "{Formation and evolution of coherent structures in 3D strongly turbulent magnetized plasmas}",
      journal = {Physics of Plasmas},
         year = 2023,
        month = apr,
       volume = {30},
       number = {4},
          eid = {040502},
        pages = {040502},
          doi = {10.1063/5.0141512},
archivePrefix = {arXiv},
       eprint = {2303.15351},
 primaryClass = {astro-ph.HE},
       adsurl = {https://ui.adsabs.harvard.edu/abs/2023PhPl...30d0502V}
}

@ARTICLE{Molinari2011,
       author = {{Molinari}, S. and {Bally}, J. and {Noriega-Crespo}, A. and {Compi{\`e}gne}, M. and {Bernard}, J.~P. and {Paradis}, D. and {Martin}, P. and {Testi}, L. and {Barlow}, M. and {Moore}, T. and {Plume}, R. and {Swinyard}, B. and {Zavagno}, A. and {Calzoletti}, L. and {Di Giorgio}, A.~M. and {Elia}, D. and {Faustini}, F. and {Natoli}, P. and {Pestalozzi}, M. and {Pezzuto}, S. and {Piacentini}, F. and {Polenta}, G. and {Polychroni}, D. and {Schisano}, E. and {Traficante}, A. and {Veneziani}, M. and {Battersby}, C. and {Burton}, M. and {Carey}, S. and {Fukui}, Y. and {Li}, J.~Z. and {Lord}, S.~D. and {Morgan}, L. and {Motte}, F. and {Schuller}, F. and {Stringfellow}, G.~S. and {Tan}, J.~C. and {Thompson}, M.~A. and {Ward-Thompson}, D. and {White}, G. and {Umana}, G.},
        title = "{A 100 pc Elliptical and Twisted Ring of Cold and Dense Molecular Clouds Revealed by Herschel Around the Galactic Center}",
      journal = {\apjl},
         year = 2011,
        month = jul,
       volume = {735},
       number = {2},
          eid = {L33},
        pages = {L33},
          doi = {10.1088/2041-8205/735/2/L33},
archivePrefix = {arXiv},
       eprint = {1105.5486},
 primaryClass = {astro-ph.GA},
       adsurl = {https://ui.adsabs.harvard.edu/abs/2011ApJ...735L..33M}
}

@ARTICLE{Yusef2022b,
       author = {{Yusef-Zadeh}, F. and {Arendt}, R.~G. and {Wardle}, M. and {Boldyrev}, S. and {Heywood}, I. and {Cotton}, W. and {Camilo}, F.},
        title = "{Statistical properties of the population of the Galactic centre filaments - II. The spacing between filaments}",
      journal = {\mnras},
         year = 2022,
        month = sep,
       volume = {515},
       number = {2},
        pages = {3059-3093},
          doi = {10.1093/mnras/stac1696},
archivePrefix = {arXiv},
       eprint = {2206.10732},
 primaryClass = {astro-ph.HE},
       adsurl = {https://ui.adsabs.harvard.edu/abs/2022MNRAS.515.3059Y}
}

@ARTICLE{Zhao2025,
       author = {{Zhao}, Roy J. and {Morris}, Mark R. and {Chuss}, David T. and {Par{\'e}}, Dylan M. and {Guerra}, Jordan A. and {Butterfield}, Natalie O. and {Wollack}, Edward J. and {Karpovich}, Kaitlyn},
        title = "{SOFIA/HAWC+ Far-infrared Polarimetric Large Area CMZ Exploration Survey. V. The Magnetic Field Strength and Morphology in the Sagittarius C Complex}",
      journal = {\apj},
         year = 2025,
        month = aug,
       volume = {988},
       number = {2},
          eid = {252},
        pages = {252},
          doi = {10.3847/1538-4357/ade80d},
archivePrefix = {arXiv},
       eprint = {2502.14961},
 primaryClass = {astro-ph.GA},
       adsurl = {https://ui.adsabs.harvard.edu/abs/2025ApJ...988..252Z}
}

@BOOK{Draine,
       author = {{Draine}, Bruce T.},
        title = "{Physics of the Interstellar and Intergalactic Medium}",
         year = 2011,
    publisher = {Princeton Series in Astrophysics},
       adsurl = {https://ui.adsabs.harvard.edu/abs/2011piim.book.....D}
}

@ARTICLE{Oka2019,
       author = {{Oka}, Takeshi and {Geballe}, T.~R. and {Goto}, Miwa and {Usuda}, Tomonori and {Benjamin} and {McCall}, J. and {Indriolo}, Nick},
        title = "{The Central 300 pc of the Galaxy Probed by Infrared Spectra of \{\{\textbackslashrm\{H\}\}\}\_\{3\}\^\{+\} and CO. I. Predominance of Warm and Diffuse Gas and High H$_{2}$ Ionization Rate}",
      journal = {\apj},
         year = 2019,
        month = sep,
       volume = {883},
       number = {1},
          eid = {54},
        pages = {54},
          doi = {10.3847/1538-4357/ab3647},
archivePrefix = {arXiv},
       eprint = {1910.04762},
 primaryClass = {astro-ph.HE},
       adsurl = {https://ui.adsabs.harvard.edu/abs/2019ApJ...883...54O}
}

@ARTICLE{Evoli2020a,
       author = {{Evoli}, Carmelo and {Morlino}, Giovanni and {Blasi}, Pasquale and {Aloisio}, Roberto},
        title = "{AMS-02 beryllium data and its implication for cosmic ray transport}",
      journal = {\prd},
         year = 2020,
        month = jan,
       volume = {101},
       number = {2},
          eid = {023013},
        pages = {023013},
          doi = {10.1103/PhysRevD.101.023013},
archivePrefix = {arXiv},
       eprint = {1910.04113},
 primaryClass = {astro-ph.HE},
       adsurl = {https://ui.adsabs.harvard.edu/abs/2020PhRvD.101b3013E}
}

@ARTICLE{Yusef2022c,
       author = {{Yusef-Zadeh}, F. and {Arendt}, R.~G. and {Wardle}, M. and {Heywood}, I. and {Cotton}, W.},
        title = "{The population of Galactic Centre filaments - III. Candidate radio and stellar sources}",
      journal = {\mnras},
         year = 2022,
        month = nov,
       volume = {517},
       number = {1},
        pages = {294-355},
          doi = {10.1093/mnras/stac2415},
archivePrefix = {arXiv},
       eprint = {2208.11589},
 primaryClass = {astro-ph.GA},
       adsurl = {https://ui.adsabs.harvard.edu/abs/2022MNRAS.517..294Y}
}

@article{Zweibel2017,
    title = {The basis for cosmic ray feedback: {Written} on the wind},
    volume = {24},
    issn = {1070-664X},
    shorttitle = {The basis for cosmic ray feedback},
    url = {https://doi.org/10.1063/1.4984017},
    doi = {10.1063/1.4984017},
    number = {5},
    urldate = {2023-09-26},
    journal = {Physics of Plasmas},
    author = {Zweibel, Ellen G.},
    month = may,
    year = {2017},
    pages = {055402},
}

@ARTICLE{Skilling1975a,
       author = {{Skilling}, J.},
        title = "{Cosmic ray streaming - I. Effect of Alfv{\'e}n waves on particles.}",
      journal = {\mnras},
         year = 1975,
        month = sep,
       volume = {172},
        pages = {557-566},
          doi = {10.1093/mnras/172.3.557},
       adsurl = {https://ui.adsabs.harvard.edu/abs/1975MNRAS.172..557S}
}

@ARTICLE{Skilling1975b,
       author = {{Skilling}, J.},
        title = "{Cosmic ray streaming - II. Effect of particles on Alfv{\'e}n waves.}",
      journal = {\mnras},
         year = 1975,
        month = nov,
       volume = {173},
        pages = {245-254},
          doi = {10.1093/mnras/173.2.245},
       adsurl = {https://ui.adsabs.harvard.edu/abs/1975MNRAS.173..245S}
}

@BOOK{Kulsrud2005,
       author = {{Kulsrud}, Russell M.},
        title = "{Plasma Physics for Astrophysics}",
    publisher = {Princeton University Press},
         year = 2005,
       adsurl = {https://ui.adsabs.harvard.edu/abs/2005ppa..book.....K}
}

@ARTICLE{Evoli2019,
       author = {{Evoli}, Carmelo and {Aloisio}, Roberto and {Blasi}, Pasquale},
        title = "{Galactic cosmic rays after the AMS-02 observations}",
      journal = {\prd},
         year = 2019,
        month = may,
       volume = {99},
       number = {10},
          eid = {103023},
        pages = {103023},
          doi = {10.1103/PhysRevD.99.103023},
archivePrefix = {arXiv},
       eprint = {1904.10220},
 primaryClass = {astro-ph.HE},
       adsurl = {https://ui.adsabs.harvard.edu/abs/2019PhRvD..99j3023E}
}

@ARTICLE{Owen2023,
       author = {{Owen}, Ellis R. and {Wu}, Kinwah and {Inoue}, Yoshiyuki and {Yang}, H.-Y. Karen and {Mitchell}, Alison M.~W.},
        title = "{Cosmic Ray Processes in Galactic Ecosystems}",
      journal = {Galaxies},
         year = 2023,
        month = jul,
       volume = {11},
       number = {4},
          eid = {86},
        pages = {86},
          doi = {10.3390/galaxies11040086},
archivePrefix = {arXiv},
       eprint = {2306.09924},
 primaryClass = {astro-ph.GA},
       adsurl = {https://ui.adsabs.harvard.edu/abs/2023Galax..11...86O}
}

@BOOK{Ginzburg1964,
       author = {{Ginzburg}, V.~L. and {Syrovatskii}, S.~I.},
        title = "{The Origin of Cosmic Rays}",
    publisher = {Pergamon Press},
         year = 1964,
       adsurl = {https://ui.adsabs.harvard.edu/abs/1964ocr..book.....G}
}

@ARTICLE{Nord2004,
       author = {{Nord}, Michael E. and {Lazio}, T. Joseph W. and {Kassim}, Namir E. and {Hyman}, S.~D. and {LaRosa}, T.~N. and {Brogan}, C.~L. and {Duric}, N.},
        title = "{High-Resolution, Wide-Field Imaging of the Galactic Center Region at 330 MHz}",
      journal = {\aj},
         year = 2004,
        month = oct,
       volume = {128},
       number = {4},
        pages = {1646-1670},
          doi = {10.1086/424001},
archivePrefix = {arXiv},
       eprint = {astro-ph/0407178},
 primaryClass = {astro-ph},
       adsurl = {https://ui.adsabs.harvard.edu/abs/2004AJ....128.1646N}
}

@ARTICLE{Aharonian2020,
       author = {{Aharonian}, Felix and {Peron}, Giada and {Yang}, Ruizhi and {Casanova}, Sabrina and {Zanin}, Roberta},
        title = "{Probing the sea of galactic cosmic rays with Fermi-LAT}",
      journal = {\prd},
         year = 2020,
        month = apr,
       volume = {101},
       number = {8},
          eid = {083018},
        pages = {083018},
          doi = {10.1103/PhysRevD.101.083018},
archivePrefix = {arXiv},
       eprint = {1811.12118},
 primaryClass = {astro-ph.HE},
       adsurl = {https://ui.adsabs.harvard.edu/abs/2020PhRvD.101h3018A}
}

@ARTICLE{Yang2023,
       author = {{Yang}, Rui-zhi and {Li}, Guang-Xing and {Wilhelmi}, Emma de O{\~n}a and {Cui}, Yu-Dong and {Liu}, Bing and {Aharonian}, Felix},
        title = "{Effective shielding of {\ensuremath{\lesssim}}10 GeV cosmic rays from dense molecular clumps}",
      journal = {Nature Astronomy},
         year = 2023,
        month = mar,
       volume = {7},
        pages = {351-358},
          doi = {10.1038/s41550-022-01868-9},
archivePrefix = {arXiv},
       eprint = {2301.06716},
 primaryClass = {astro-ph.HE},
       adsurl = {https://ui.adsabs.harvard.edu/abs/2023NatAs...7..351Y}
}

@ARTICLE{Abeysekara2017,
       author = {{Abeysekara}, A.~U. and {Albert}, A. and {Alfaro}, R. and {Alvarez}, C. and {{\'A}lvarez}, J.~D. and {Arceo}, R. and {Arteaga-Vel{\'a}zquez}, J.~C. and {Avila Rojas}, D. and {Ayala Solares}, H.~A. and {Barber}, A.~S. and {Bautista-Elivar}, N. and {Becerril}, A. and {Belmont-Moreno}, E. and {BenZvi}, S.~Y. and {Berley}, D. and {Bernal}, A. and {Braun}, J. and {Brisbois}, C. and {Caballero-Mora}, K.~S. and {Capistr{\'a}n}, T. and {Carrami{\~n}ana}, A. and {Casanova}, S. and {Castillo}, M. and {Cotti}, U. and {Cotzomi}, J. and {Couti{\~n}o de Le{\'o}n}, S. and {De Le{\'o}n}, C. and {De la Fuente}, E. and {Dingus}, B.~L. and {DuVernois}, M.~A. and {D{\'\i}az-V{\'e}lez}, J.~C. and {Ellsworth}, R.~W. and {Engel}, K. and {Enr{\'\i}quez-Rivera}, O. and {Fiorino}, D.~W. and {Fraija}, N. and {Garc{\'\i}a-Gonz{\'a}lez}, J.~A. and {Garfias}, F. and {Gerhardt}, M. and {Gonz{\'a}lez Mu{\~n}oz}, A. and {Gonz{\'a}lez}, M.~M. and {Goodman}, J.~A. and {Hampel-Arias}, Z. and {Harding}, J.~P. and {Hern{\'a}ndez}, S. and {Hern{\'a}ndez-Almada}, A. and {Hinton}, J. and {Hona}, B. and {Hui}, C.~M. and {H{\"u}ntemeyer}, P. and {Iriarte}, A. and {Jardin-Blicq}, A. and {Joshi}, V. and {Kaufmann}, S. and {Kieda}, D. and {Lara}, A. and {Lauer}, R.~J. and {Lee}, W.~H. and {Lennarz}, D. and {Vargas}, H. Le{\'o}n and {Linnemann}, J.~T. and {Longinotti}, A.~L. and {Luis Raya}, G. and {Luna-Garc{\'\i}a}, R. and {L{\'o}pez-Coto}, R. and {Malone}, K. and {Marinelli}, S.~S. and {Martinez}, O. and {Martinez-Castellanos}, I. and {Mart{\'\i}nez-Castro}, J. and {Mart{\'\i}nez-Huerta}, H. and {Matthews}, J.~A. and {Miranda-Romagnoli}, P. and {Moreno}, E. and {Mostaf{\'a}}, M. and {Nellen}, L. and {Newbold}, M. and {Nisa}, M.~U. and {Noriega-Papaqui}, R. and {Pelayo}, R. and {Pretz}, J. and {P{\'e}rez-P{\'e}rez}, E.~G. and {Ren}, Z. and {Rho}, C.~D. and {Rivi{\`e}re}, C. and {Rosa-Gonz{\'a}lez}, D. and {Rosenberg}, M. and {Ruiz-Velasco}, E. and {Salazar}, H. and {Salesa Greus}, F. and {Sandoval}, A. and {Schneider}, M. and {Schoorlemmer}, H. and {Sinnis}, G. and {Smith}, A.~J. and {Springer}, R.~W. and {Surajbali}, P. and {Taboada}, I. and {Tibolla}, O. and {Tollefson}, K. and {Torres}, I. and {Ukwatta}, T.~N. and {Vianello}, G. and {Weisgarber}, T. and {Westerhoff}, S. and {Wisher}, I.~G. and {Wood}, J. and {Yapici}, T. and {Yodh}, G. and {Younk}, P.~W. and {Zepeda}, A. and {Zhou}, H. and {Guo}, F. and {Hahn}, J. and {Li}, H. and {Zhang}, H.},
        title = "{Extended gamma-ray sources around pulsars constrain the origin of the positron flux at Earth}",
      journal = {Science},
         year = 2017,
        month = nov,
       volume = {358},
       number = {6365},
        pages = {911-914},
          doi = {10.1126/science.aan4880},
archivePrefix = {arXiv},
       eprint = {1711.06223},
 primaryClass = {astro-ph.HE},
       adsurl = {https://ui.adsabs.harvard.edu/abs/2017Sci...358..911A}
}

@ARTICLE{Ng2026,
       author = {{Ng}, Hayden P.~H. and {Owen}, Ellis R. and {Tsuji}, Naomi and {Chen}, Szu-Ting},
        title = "{Multiwavelength probes of cosmic ray transport in molecular cloud structures}",
      journal = {\pasj},
         year = 2026,
        month = aug,
       volume = {78},
       number = {4},
        pages = {1390-1410},
          doi = {10.1093/pasj/psag070},
archivePrefix = {arXiv},
       eprint = {2605.16698},
 primaryClass = {astro-ph.HE},
       adsurl = {https://ui.adsabs.harvard.edu/abs/2026PASJ...78.1390N}
}

@ARTICLE{YusefZadeh2003,
       author = {{Yusef-Zadeh}, F.},
        title = "{The Origin of the Galactic Center Nonthermal Radio Filaments: Young Stellar Clusters}",
      journal = {\apj},
         year = 2003,
        month = nov,
       volume = {598},
       number = {1},
        pages = {325-333},
          doi = {10.1086/378715},
archivePrefix = {arXiv},
       eprint = {astro-ph/0308008},
 primaryClass = {astro-ph},
       adsurl = {https://ui.adsabs.harvard.edu/abs/2003ApJ...598..325Y}
}

@ARTICLE{Dahlburg2002,
       author = {{Dahlburg}, Russell B. and {Einaudi}, Giorgio and {LaRosa}, T.~N. and {Shore}, Steven N.},
        title = "{Modeling the Galactic Center Nonthermal Filaments as Magnetized Wakes}",
      journal = {\apj},
         year = 2002,
        month = mar,
       volume = {568},
       number = {1},
        pages = {220-225},
          doi = {10.1086/338842},
archivePrefix = {arXiv},
       eprint = {astro-ph/0112378},
 primaryClass = {astro-ph},
       adsurl = {https://ui.adsabs.harvard.edu/abs/2002ApJ...568..220D}
}

@ARTICLE{Barkov2019,
       author = {{Barkov}, Maxim V. and {Lyutikov}, Maxim},
        title = "{On the nature of radio filaments near the Galactic Centre}",
      journal = {\mnras},
         year = 2019,
        month = oct,
       volume = {489},
       number = {1},
        pages = {L28-L31},
          doi = {10.1093/mnrasl/slz124},
archivePrefix = {arXiv},
       eprint = {1905.13103},
 primaryClass = {astro-ph.HE},
       adsurl = {https://ui.adsabs.harvard.edu/abs/2019MNRAS.489L..28B}
}

@ARTICLE{Bicknell2001,
       author = {{Bicknell}, Geoffrey V. and {Li}, Jianke},
        title = "{The Snake: A Reconnecting Coil in a Twisted Magnetic Flux Tube}",
      journal = {\apjl},
         year = 2001,
        month = feb,
       volume = {548},
       number = {1},
        pages = {L69-L72},
          doi = {10.1086/318928},
archivePrefix = {arXiv},
       eprint = {astro-ph/0011382},
 primaryClass = {astro-ph},
       adsurl = {https://ui.adsabs.harvard.edu/abs/2001ApJ...548L..69B}
}

@ARTICLE{Linden2011,
       author = {{Linden}, Tim and {Hooper}, Dan and {Yusef-Zadeh}, Farhad},
        title = "{Dark Matter and Synchrotron Emission from Galactic Center Radio Filaments}",
      journal = {\apj},
         year = 2011,
        month = nov,
       volume = {741},
       number = {2},
          eid = {95},
        pages = {95},
          doi = {10.1088/0004-637X/741/2/95},
archivePrefix = {arXiv},
       eprint = {1106.5493},
 primaryClass = {astro-ph.HE},
       adsurl = {https://ui.adsabs.harvard.edu/abs/2011ApJ...741...95L}
}

@ARTICLE{Ewart2026,
       author = {{Ewart}, Robert J. and {Reichherzer}, Patrick and {Ren}, Shuzhe and {Majeski}, Stephen and {Mori}, Francesco and {Nastac}, Michael L. and {Bott}, Archie F.~A. and {Kunz}, Matthew W. and {Schekochihin}, Alexander A.},
        title = "{Cosmic-ray transport in inhomogeneous media}",
      journal = {\mnras},
         year = 2026,
        month = jan,
       volume = {545},
       number = {2},
          eid = {staf2108},
        pages = {staf2108},
          doi = {10.1093/mnras/staf2108},
archivePrefix = {arXiv},
       eprint = {2507.19044},
 primaryClass = {astro-ph.HE},
       adsurl = {https://ui.adsabs.harvard.edu/abs/2026MNRAS.545f2108E}
}

@ARTICLE{Fulat2023,
       author = {{Fulat}, Karol and {Bohdan}, Artem and {Torralba Paz}, Gabriel and {Pohl}, Martin},
        title = "{Kinetic Simulations of Nonrelativistic High-mach-number Perpendicular Shocks Propagating in a Turbulent Medium}",
      journal = {\apj},
         year = 2023,
        month = dec,
       volume = {959},
       number = {2},
          eid = {119},
        pages = {119},
          doi = {10.3847/1538-4357/ad04dc},
archivePrefix = {arXiv},
       eprint = {2310.14388},
 primaryClass = {physics.plasm-ph},
       adsurl = {https://ui.adsabs.harvard.edu/abs/2023ApJ...959..119F}
}

@ARTICLE{Fulat2026,
       author = {{Fulat}, Karol and {Moore}, Eloise and {Alawashra}, Mahmoud and {Tsirou}, Michelle and {Bohdan}, Artem and {Amano}, Takanobu and {Pohl}, Martin},
        title = "{PIC Simulations of Nonrelativistic High-Mach-number Oblique Shocks Propagating in a Turbulent Medium}",
      journal = {\apj},
         year = 2026,
        month = jul,
       volume = {1005},
       number = {1},
          eid = {107},
        pages = {107},
          doi = {10.3847/1538-4357/ae71b6},
archivePrefix = {arXiv},
       eprint = {2605.21986},
 primaryClass = {physics.plasm-ph},
       adsurl = {https://ui.adsabs.harvard.edu/abs/2026ApJ..1005..107F}
}

@ARTICLE{Federrath2016,
       author = {{Federrath}, C. and {Rathborne}, J.~M. and {Longmore}, S.~N. and {Kruijssen}, J.~M.~D. and {Bally}, J. and {Contreras}, Y. and {Crocker}, R.~M. and {Garay}, G. and {Jackson}, J.~M. and {Testi}, L. and {Walsh}, A.~J.},
        title = "{The Link between Turbulence, Magnetic Fields, Filaments, and Star Formation in the Central Molecular Zone Cloud G0.253+0.016}",
      journal = {\apj},
         year = 2016,
        month = dec,
       volume = {832},
       number = {2},
          eid = {143},
        pages = {143},
          doi = {10.3847/0004-637X/832/2/143},
archivePrefix = {arXiv},
       eprint = {1609.05911},
 primaryClass = {astro-ph.GA},
       adsurl = {https://ui.adsabs.harvard.edu/abs/2016ApJ...832..143F}
}

@ARTICLE{Hooper2018,
       author = {{Hooper}, Dan and {Cholis}, Ilias and {Linden}, Tim},
        title = "{TeV gamma rays from Galactic Center pulsars}",
      journal = {Physics of the Dark Universe},
         year = 2018,
        month = sep,
       volume = {21},
          eid = {40},
        pages = {40},
          doi = {10.1016/j.dark.2018.05.004},
archivePrefix = {arXiv},
       eprint = {1705.09293},
 primaryClass = {astro-ph.HE},
       adsurl = {https://ui.adsabs.harvard.edu/abs/2018PDU....21...40H}
}

@ARTICLE{Desiati2014,
       author = {{Desiati}, Paolo and {Zweibel}, Ellen G.},
        title = "{The Transport of Cosmic Rays Across Magnetic Fieldlines}",
      journal = {\apj},
         year = 2014,
        month = aug,
       volume = {791},
       number = {1},
          eid = {51},
        pages = {51},
          doi = {10.1088/0004-637X/791/1/51},
archivePrefix = {arXiv},
       eprint = {1402.1475},
 primaryClass = {astro-ph.HE},
       adsurl = {https://ui.adsabs.harvard.edu/abs/2014ApJ...791...51D}
}

@ARTICLE{Koldobskiy2021,
       author = {{Koldobskiy}, S. and {Kachelrie{\ss}}, M. and {Lskavyan}, A. and {Neronov}, A. and {Ostapchenko}, S. and {Semikoz}, D.~V.},
        title = "{Energy spectra of secondaries in proton-proton interactions}",
      journal = {\prd},
         year = 2021,
        month = dec,
       volume = {104},
       number = {12},
          eid = {123027},
        pages = {123027},
          doi = {10.1103/PhysRevD.104.123027},
archivePrefix = {arXiv},
       eprint = {2110.00496},
 primaryClass = {astro-ph.HE},
       adsurl = {https://ui.adsabs.harvard.edu/abs/2021PhRvD.104l3027K}
}
\bibliographystyle{aasjournal}

\appendix

\section{Convergence Tests \label{sec:convergence}}

\begin{figure}[h]
    \centering
    \includegraphics[width=\linewidth]{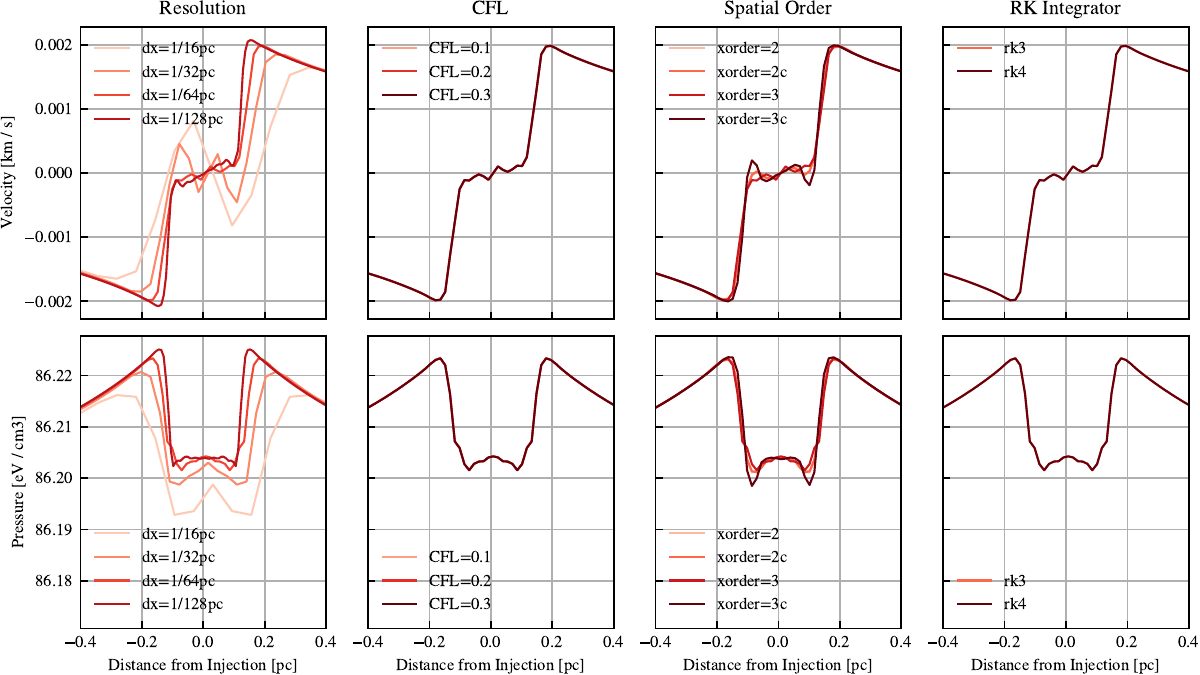}
    \caption{Plots of the velocity and pressure 10 kyr after a lepton-only injection, zoomed in on the center pc of the simulated filament. When not varying a specific parameter, these convergence tests were run with a resolution of $dx=1/64 \pc$, a CFL number of 0.2, a spatial order of 3, and using the RK integrator ``rk4.'' For the complete suite of simulations, we set $dx=1/64 \pc$ in order to minimize the jagged velocity between $\pm0.1 \pc$ as well as flatten the miniature peak in pressure at the center.}
    \label{fig:convergence_all}
\end{figure}

Our full suite of simulations used a Courant-Friedrich-Lewy (CFL) number of 0.2, a resolution of $dx=1/64 \pc$, a spatial order of 3, and using the Runge-Kutta (RK) time integrator ``rk4.'' These were chosen by running multiple simulations where we vary each choice and determining which values were the most accurate while also being computationally inexpensive. The data we used to make this choice are plotted in Figure \ref{fig:convergence_all}.  We find that the choice of CFL number of RK time integrator did not affect the fidelity of the simulation, while the spatial order and resolution did. Most notably, the resolution had potential discontinuous, jagged effects if the resolution were less than $dx=1/64\pc$. While doubling the resolution to $dx=1/128\pc$ made the curves slightly smoother, it did not notably change the results, so we kept $dx=1/64\pc$ for computational cheapness.

\section{Auxiliary Calculations}

\subsection{Adiabatic Expansion of Flux Tube \label{sec:adiabatic}}

To justify why we are allowed to do a 1-dimensional simulation, we aim to show here that by an adiabatic injection of \textmra{CR} energy $\Delta U_c$ into a flux tube, the tube should only change size perturbatively to remain in pressure balance with its surroundings. This will also quantify the conclusion in Section \ref{sec:heating} that \textmra{CR} injection is unlikely to drive a shock.

The plasma $\beta\equiv 8\pi P_g/B^2\sim 0.17$ for fiducial parameters is low enough that, in estimating the fractional change in tube radius $\Delta R/R$, we err on the side of ignoring gas pressure and simply assume that the addition of \textmra{CR} pressure $\Delta P_c=\Delta U_c/3$ is compensated by a reduction in magnetic pressure:
\begin{equation}
\Delta\frac{B^2}{8\pi}=-\Delta P_c\sim\frac{B\Delta B}{4\pi}\,.
\end{equation}

By magnetic flux conservation, $(B+\Delta B)(R+\Delta R)^2=BR^2$, so to first order in the perturbation amplitude,
\begin{equation}
\frac{\Delta R}{R}\sim -\frac{\Delta B}{2B}.
\end{equation}

Combining the pressure balance and flux conservation relations gives
\begin{equation}
\frac{\Delta R}{R}\sim\frac{2\pi\Delta P_c}{B^2}.
\end{equation}

We estimate $\Delta P_c$ in several different ways. To put an upper bound on the size of the effect, we assume that $\Delta U_c=5\times 10^{42}\times 100$ ergs (i.e. the fiducial injected \textmra{CR} lepton energy multiplied by 100 to account for \textmra{CR} protons) is released instantaneously into a sphere of radius 0.1 pc. This gives $\Delta R/R\sim 0.21$. Accounting for propagation over our fiducial injection time of 2 kyr  at our fiducial $\kappa=3\times 10^{26} \cmcms$ means the volume would be $\sim 20$ times larger, reducing $\Delta R/R$ to $\sim 2$\%. If protons were absent, both estimates of $\Delta R/R$ would be reduced by 2 orders of magnitude. We conclude that even the worst case scenario - instantaneous injection of both leptons and protons, ignoring the role of thermal gas pressure, and neglecting propagation - results in only a 20\% effect which would be significantly reduced by a number of effects. By the same token, the overpressure resulting from \textmra{CR} injection is probably not large enough to drive a strong shock.

\subsection{Luminosity Derivation \label{sec:luminosityderivation}}

Here, we derive a functional formula for the luminosity emitted from an NTF for both the purposes of setting parameters in Table \ref{tab:Simulations_run} and for creating luminosity plots as in Figure \ref{fig:EcLc}. We begin by noting relations between $\nu$ and $\gamma$:
\begin{equation}
    \nu = \frac{3}{16}\frac{eB}{m_ec}\gamma^2; \quad
    \frac{d\nu}{d\gamma} = \frac{3}{8}\frac{eB}{m_ec}\gamma = \frac{2\nu}{\gamma}\,.
\end{equation}

Assuming that synchrotron radiation is the predominant source, a single electron will experience energy losses of
\begin{equation}
    m_ec^2\dot\gamma = \mathmra{\dot{E_c}}(\gamma) = -\frac{1}{6\pi}\sigma_Tc\gamma^2B^2\,.
\end{equation}

Integrating over an entire spectrum of particles, $f(\gamma) = C\gamma^{-\alpha}$,
\begin{equation}
    L_\gamma = -\frac{1}{6\pi}\sigma_Tc\gamma^2B^2f(\gamma)\,.
\end{equation}

Lastly, we need to convert $L_\gamma$ to $L_\nu$:
\begin{equation}
    L_\nu = L_\gamma\frac{d\gamma}{d\nu} = L_\gamma\frac{\gamma}{2\nu} = \frac{L_\gamma}{2\nu}\sqrt{\frac{16m_ec}{3e}\frac{\nu}{B}}\,,
\end{equation}
\begin{equation}
    L_\nu = \frac{\sigma_Tc\gamma^3B^2}{12\pi\nu}f(\gamma) = \frac{C\sigma_Tc\gamma^{3-\alpha}B^2}{12\pi\nu} = \frac{C\sigma_TcB^2}{12\pi\nu}\left(\frac{16m_ec}{3e}\frac{\nu}{B}\right)^{(3-\alpha)/2}\,,
\end{equation}
\begin{equation}
    L_\nu=\frac{C\sigma_Tc}{12\pi}\left(\frac{16m_ec}{3e}\right)^{(3-\alpha)/2}\nu^{(1-\alpha)/2}B^{(1+\alpha)/2}\,.
\end{equation}

The three variables that matter here are $C$, $B$, and $\nu$. When we keep both $L_\nu$ and $\nu$, we must adjust $C$, a factor proportional to the total injection energy, to change with the magnetic field $B$. In other words, stronger magnetic fields require less \textmra{CRs} to maintain the same emission strength, as $C\propto B^{-(1+\alpha)/2}.$

In order to plot this luminosity $L_\nu$, we must acknowledge the background CR energy density $\mathmra{\varepsilon_{c,0}}=0.01\eVcm$ omnipresent in the ISM (which we assume shares a spectral index with the injection). Therefore, we obtain that
\begin{equation}
    L_\nu = A * (\mathmra{\varepsilon_c} + \mathmra{\varepsilon_{c,0}}) * B^{(1+\alpha)/2}\,,
\end{equation}
\begin{equation}\label{eq:luminosity}
    \frac{L_\nu}{A\mathmra{\varepsilon_{c,0}}B_0^{(1+\alpha)/2}} = \left(\frac{\mathmra{\varepsilon_c}}{\mathmra{\varepsilon_{c,0}}}+1\right)\left(\frac{B}{B_0}\right)^{(1+\alpha)/2}\,.
\end{equation}

This equation is set up such that the right-hand side equals 1 when the emissivity of the filament is the same as if there were no injection at fiducial parameters. Any number other than 1 is the enhancement of the luminosity above this background (e.g. 2 would mean its emissivity is twice as strong).

\subsection{Self-Confinement Model\label{sec:selfconfinement}}
Here we assess the plausibility of \textmra{CR} self-confinement, with and without proton \textmra{CRs}. We assume that the \textmra{CR} drift $v_D$ is regulated by balancing \Alfven wave excitation by the streaming instability against damping, and that nonlinear Landau damping is the main damping mechanism:%
\begin{equation}\label{eq:marginalstability}
\frac{\pi}{4}\frac{\alpha - 1}{\alpha}\omega_{ci}\frac{n_{cr}(>p_1)}{n_i}\left(\frac{v_D}{v_A}-1  \right)=\left(\frac{\pi}{32} \right)^{1/2}kv_i\left(\frac{\delta B}{B} \right)^2,
\end{equation}
where $\alpha$ is the \textmra{CR} energy spectral index, $k$ is the wavenumber of the wave, $v_i\equiv\sqrt{k_BT/m_i}$, $\delta B/B$ is dimensionless wave amplitude, and $p_1\equiv eB/kc$ is the minimum momentum that can be gyroresonant with the wave. 

A second relation comes from the relationship between the anisotropy and the scattering rate $\nu$, and \textmra{CR} pressure gradient lengthscale $L$,
\begin{equation}\label{eq:anisotropy}
\frac{v_D-v_A}{c}\sim\frac{c}{\nu L}\,.
\end{equation}
The scattering rate of a particle with gyrofrequency $\omega_{cr}$ is related to the wave amplitude by
\begin{equation}\label{eq:nu}
\nu\sim\frac{\pi}{8}\omega_{cr} \left(\frac{\delta B}{B} \right)^2\,.
\end{equation}
Relations (\ref{eq:marginalstability}--\ref{eq:nu}) can be found within \cite{Kulsrud2005}). Combining them gives
\begin{equation}\label{eq:driftsqrd}
\left(\frac{v_D}{v_A} -1  \right)^2 = \left(\frac{32}{\pi^3} \right)^{3/2}\frac{r_i}{L}\frac{c}{v_A}\frac{n_i}{n_{cr}}\frac{\alpha}{\alpha - 1}\,.
\end{equation}
In Equation (\ref{eq:driftsqrd}), $r_i\equiv v_i/\omega_{ci}$ and $n_{cr}$ is the total injected \textmra{CR} density. For fiducial parameters and assuming $5\times 10^{42}$ ergs in $\gamma = 3500$ $e^{\pm}$ in a cylinder of length 20 pc and radius 0.1 pc we find
\begin{equation}\label{eq:drift}
v_D/v_A =10.5\,.
\end{equation} 
Using Equation (\ref{eq:drift}) in Equations (\ref{eq:marginalstability})\&(\ref{eq:nu}), we find
\begin{equation}\label{eq:kappa}
\kappa=\frac{c^2}{\nu}=2.3\times 10^{27}\textrm{cm}^2\textrm{s}^{-1}\,.   
\end{equation}
Equations (\ref{eq:drift}) and (\ref{eq:kappa}) suggest that self-confinement does not work for the relativistic leptons alone. Interestingly, adding protons \textit{does} produce a satisfactory model; $v_D/v_A - 1$ is reduced by a factor of 10, and  $\kappa$ is decreased by a factor of 100.

\end{document}